\documentclass[12pt]{iopart}

\expandafter\let\csname equation*\endcsname=\relax 
\expandafter\let\csname endequation*\endcsname=\relax 
\usepackage{amsmath,amsfonts}

\usepackage[
backend=biber,
style=numeric-comp,
sorting=none,
mincitenames=1,
maxcitenames=2,
]{biblatex} 
\usepackage{graphicx}
\usepackage{bm}

\usepackage[dvipsnames]{xcolor}

\usepackage{siunitx}[=v2]
\usepackage{xspace}
\usepackage[acronym]{glossaries}
\glsdisablehyper

\usepackage{booktabs}
\usepackage{subfigure}
\usepackage{enumitem}

\usepackage{hyperref}
\usepackage{cleveref}
\crefname{appendix}{}{}

\usepackage[linesnumbered,ruled]{algorithm2e}
\SetKwInput{KwInput}{Input}                
\SetKwInput{KwOutput}{Output}              
\crefname{algocf}{alg.}{algs.}
\Crefname{algocf}{Algorithm}{Algorithms}

\begin{document}

\newcommand{\nball}[1]{$#1$\nobreakdash\discretionary{-}{-}{-}ball }
\newcommand{\nsphere}[1]{$#1$\nobreakdash\discretionary{-}{-}{-}sphere }
\newcommand{\ndimensional}[1]{$#1$\nobreakdash\discretionary{-}{-}{-}dimensional}
\newcommand{\latent}{\mathcal{Z}}
\newcommand{\physical}{\mathcal{X}}
\newcommand{\diff}{\textrm{d}}
\renewcommand{\vec}[1]{\boldsymbol{#1}}
\newcommand{\nlive}{N_{\textrm{live}}}
\newcommand{\evidence}{Z}
\newcommand{\likelihood}{\mathcal{L}}
\newcommand{\prior}{\pi}
\newcommand{\likelihoodthreshold}{\likelihood_{t}}

\newcommand{\kl}{\textrm{KL}}

\newcommand{\codestyle}[1]{\texttt{#1}}

\newcommand{\nessai}{\codestyle{nessai}\xspace}
\newcommand{\inessai}{\codestyle{i\protect\nobreakdash-nessai}\xspace}
\newcommand{\bilby}{\codestyle{bilby}\xspace}
\newcommand{\bilbypipe}{\codestyle{bilby\_pipe}\xspace}
\newcommand{\lalinference}{\codestyle{LALInference}\xspace}
\newcommand{\lalsuite}{\codestyle{LALSuite}}
\newcommand{\dynesty}{\codestyle{dynesty}\xspace}
\newcommand{\cpnest}{\codestyle{cpnest}\xspace}
\newcommand{\multinest}{\codestyle{multinest}\xspace}
\newcommand{\nflows}{\codestyle{nflows}\xspace}
\newcommand{\glasflow}{\codestyle{glasflow}\xspace}
\newcommand{\nessaimodels}{\codestyle{nessai-models}\xspace}
\newcommand{\pytorch}{\codestyle{PyTorch}\xspace}
\newcommand{\corner}{\codestyle{corner}\xspace}
\newcommand{\matplotlib}{\codestyle{matplotlib}\xspace}
\newcommand{\seaborn}{\codestyle{seaborn}\xspace}
\newcommand{\numpy}{\codestyle{NumPy}\xspace}
\newcommand{\scipy}{\codestyle{SciPy}\xspace}
\newcommand{\pandas}{\codestyle{pandas}\xspace}
\newcommand{\python}{\codestyle{Python}\xspace}
\newcommand{\statsmodels}{\codestyle{statsmodels}\xspace}

\newcommand{\imrphenomp}{\codestyle{IMRPhenomPv2}\xspace}
\newcommand{\imrphenomptidal}{\codestyle{IMRPhenomPv2\_NRtidalv2}\xspace}

\newcommand{\tocite}[1]{[cite \textbf{#1}]}

\newcommand{\ntotal}{N_\textrm{Total}}
\newcommand{\nperflow}{N_{j}}
\newcommand{\nremoved}{M_{j}}

\newcommand{\toyexamplepriorstd}{\checked{2}}
\newcommand{\toyexamplendims}{\checked{2}}
\newcommand{\toyexamplendist}{\checked{4}}
\newcommand{\toyexampleevidence}{\checked{0.03177}}
\newcommand{\toyexampleevidenceerror}{\checked{0.00042}}
\newcommand{\toyexampleevidencefinal}{\checked{0.03191}}
\newcommand{\toyexampleevidenceerrorfinal}{\checked{0.00042}}
\newcommand{\toyexampleanalyticevidence}{\checked{0.03183}}

\newcommand{\nanalytictests}{\checked{50}\xspace}
\newcommand{\averagebias}{\checked{0.6}\%\xspace}
\newcommand{\averagebiasgmm}{\checked{0.9}\%\xspace}

\newcommand{\nlivetimeincrease}{\checked{\sim22}\xspace}

\newcommand{\bnssamplingfrequency}{8192}
\newcommand{\bnsduration}{\checked{80}\xspace}
\newcommand{\bnsdistance}{45}
\newcommand{\bnsnetworksnr}{\checked{30.12}\xspace}
\newcommand{\jsdsamples}{\checked{5,000} \xspace}
\newcommand{\inessaibnsevals}{\checked{$1.01 \times 10^{6}$}\xspace}
\newcommand{\inessaibnstime}{\checked{24}\xspace}
\newcommand{\nessaibnsevals}{\checked{$1.42 \times 10^{6}$}\xspace}
\newcommand{\nessaibnstime}{\checked{58}\xspace}
\newcommand{\dynestybnsevals}{\checked{$4.30 \times 10^{7}$}\xspace}
\newcommand{\dynestybnstime}{\checked{376}\xspace}
\newcommand{\inessaievalsnessai}{\checked{1.4}\xspace}
\newcommand{\inessaibnsspeedup}{\checked{2.4}\xspace}
\newcommand{\inessaievalsdynesty}{\checked{42.5}\xspace}
\newcommand{\inessaibnsspeedupdynesty}{\checked{15.5}\xspace}

\newcommand{\bbhsamplingfrequency}{2048}
\newcommand{\bbhduration}{\checked{4}\xspace}
\newcommand{\numbbhsignals}{\checked{64}\xspace}
\newcommand{\bbhmedianlikelihoodevals}{\checked{$6.5\times10^{5}$}\xspace}
\newcommand{\bbhmediantime}{\checked{119} minutes\xspace}
\newcommand{\bbhlikelihoodratio}{\checked{2.68}\xspace}
\newcommand{\bbhtimeratio}{\checked{4.2}\xspace}
\newcommand{\bbhlikelihoodratiodynesty}{\checked{13.3}\xspace}
\newcommand{\bbhtimeratiodynesty}{\checked{17.2}\xspace}
\newcommand{\combinedpvalue}{\checked{0.3798}\xspace}

\newcommand{\nessaipopulationtime}{\checked{40\%}\xspace}
\newcommand{\nessaitrainingtime}{\checked{8\%}\xspace}
\newcommand{\inessaiothertime}{\checked{13\%}\xspace}

\newacronym{ess}{ESS}{effective sample size}

\newacronym{snr}{SNR}{signal-to-noise ratio}

\newacronym{roq}{ROQ}{Reduced-Order-Quadrature}

\newacronym{lvk}{LVK}{LIGO-Virgo-KAGRA Collaboration}

\newacronym{mcmc}{MCMC}{Markov Chain Monte Carlo}
\newacronym{lfi}{LFI}{likelihood-free inference}
\newacronym{mlp}{MLP}{multilayer perceptron}

\newacronym{pptest}{P-P test}{probability-probability test}

\newacronym{smc}{SMC}{Sequential Monte Carlo}
\newacronym{ns-smc}{NS-SMC}{Nested Sampling via Sequential Monte Carlo}
\newacronym{iid}{i.i.d.}{independently and identically distributed}
\newacronym{lars}{LARS}{learned acceptance/rejection sampling}
\newacronym{kld}{KL divergence}{Kullback-Leibler divergence}
\newacronym{jsd}{JS divergence}{Jensen-Shannon divergence}

\newcommand{\michael}[1]{{{\color{blue}Michael: #1}}}
\newcommand{\chris}[1]{{{\color{orange}Chris: #1}}}
\newcommand{\jv}[1]{{\color{Green}#1}}

\newcommand{\figwidth}{5.18158in}
\newcommand{\halffigwidth}{2.59079in}

\newcommand{\checkme}[1]{{\color{red} #1}}
\newcommand{\checked}[1]{{\color{blue} #1}}
\renewcommand{\checked}[1]{#1}

\newcommand{\qty}[2]{\SI{#1}{#2}}

\title{Importance nested sampling with normalising flows}
\author{Michael J. Williams, John Veitch, Chris Messenger}
\address{SUPA, School of Physics and Astronomy, University of Glasgow, Glasgow G12 8QQ, United Kingdom}
\ead{m.williams.4@research.gla.ac.uk}

\begin{abstract}
    We present an improved version of the nested sampling algorithm \nessai in which the core algorithm is modified to use importance weights. In the modified algorithm, samples are drawn from a mixture of normalising flows and the requirement for samples to be \glsfirst{iid} according to the prior is relaxed. Furthermore, it allows for samples to be added in any order, independently of a likelihood constraint, and for the evidence to be updated with batches of samples. We call the modified algorithm \inessai. We first validate \inessai using analytic likelihoods with known Bayesian evidences and show that the evidence estimates are unbiased in up to 32 dimensions. We compare \inessai to standard \nessai for the analytic likelihoods and the Rosenbrock likelihood, the results show that \inessai is consistent with \nessai whilst producing more precise evidence estimates. We then test \inessai on \numbbhsignals simulated gravitational-wave signals from binary black hole coalescence and show that it produces unbiased estimates of the parameters. We compare our results to those obtained using standard \nessai and \dynesty and find that \inessai requires \bbhlikelihoodratio and \bbhlikelihoodratiodynesty times fewer likelihood evaluations to converge, respectively. We also test \inessai of an \bnsduration-second simulated binary neutron star signal using a \glsfirst{roq} basis and find that, on average, it converges in \inessaibnstime minutes, whilst only requiring \inessaibnsevals likelihood evaluations compared to \nessaibnsevals for \nessai and \dynestybnsevals for \dynesty. These results demonstrate that \inessai is consistent with \nessai and \dynesty whilst also being more efficient.
\end{abstract}


\submitto{Machine Learning: Science and Technology}

\maketitle

\glsresetall

\section{Introduction}

 John Skilling proposed nested sampling in \cite{nested_sampling,nested_sampling_1} and it has since seen widespread use in astronomical data analysis, including but not limited to the analyses of gravitational waves \cite{lalinference,bilby}, asteroseismology \cite{diamonds} and cosmology \cite{polychord}.

Nested sampling is a Monte Carlo algorithm that approximates the Bayesian evidence
\begin{equation}
    Z \equiv p(d|H) = \int p(d| \vec{\theta}, H) \diff\vec{\theta},
\end{equation}
for some observed data $d$ with an assumed model $H$ over the parameters $\vec{\theta}$ where $\likelihood(\vec{\theta})\equiv p(d| \vec{\theta}, H)$ is the likelihood. This is usually considered in the context of Bayes' theorem
\begin{equation}\label{eq:bayes_theorem}
    p(\vec{\theta}|d, H) = \frac{p(d|\vec{\theta}, H)p(\vec{\theta}|H)}{p(d|H)},
\end{equation}
where $\pi(\vec{\theta}) \equiv p(\vec{\theta}|H)$ is the prior and $p(\vec{\theta}|d, H)$ is the posterior. Samples from the latter are a by-product of approximating the evidence.

When implementing nested sampling, the main challenge is drawing new points from the likelihood-constrained prior at a given iteration. There are different approaches to this such as using \gls{mcmc}, slice sampling or sampling from bounding distributions \cite{buchner_review}. 
There have also been efforts to incorporate machine learning into nested sampling for approximating the likelihood~\cite{bambi}, in the proposal process~\cite{moss-ns-flows, nessai} and for sampling from arbitrary priors~\cite{Alsing:2021wef}.

In \textcite{nessai}, we proposed \nessai, a nested sampling algorithm that uses normalising flows to approximate the likelihood-constrained prior at different iterations. We showed that this approach could speed up convergence and allowed for natural parallelisation of the likelihood.
However, we noted that a significant portion of compute time was being spent performing rejection sampling to ensure points were distributed according to the prior, and this, alongside the inherently serial nature of nested sampling, set a lower limit on how fast the algorithm could be.

In this work, we present a modified nested sampling algorithm based on importance sampling that addresses the aforementioned bottlenecks. In particular, this modified algorithm:
\begin{itemize}
    \item incorporates normalising flows in a similar fashion to \citeauthor{nessai}~\cite{nessai},
    \item removes the requirement for samples to be \gls{iid} and distributed according to the prior,
    \item allows samples to be added in any order independent of a likelihood constraint,
    \item allows the evidence to be updated for batches of samples.
\end{itemize}
Taken together, these changes improve the efficiency of the algorithm, reducing the number of required likelihood evaluations by up to an order of magnitude over our previous version, and greatly increasing the scalability of the algorithm. 

This is especially relevant in the context of gravitational-wave data analysis, where nested sampling is the de facto analysis algorithm~\cite{lalinference,bilby}. As of the last LIGO-Virgo-KAGRA~\cite{LIGO,Virgo,KAGRA} observing run, there are 90 confirmed detected compact binaries~\cite{gwtc2,gwtc2.1,gwtc3} and this number is expected to increase by a factor of $\sim3.3$ in the fourth observing run~\cite{observing_scenarios}. This presents a significant computational challenge since typical analyses take of order days to weeks. Furthermore, a subset of these analyses are currently only possible at great computational cost \cite{pbilby,rift}. The algorithm we present brings the possibility of tackling these challenging analyses and dramatically reduces the wall-time required to complete an analysis.

This paper is structured as follows: in \cref{sec:background} we present background theory on nested sampling and various alternative formulations that this work builds upon. We then describe a simplified version of our modified algorithm and validate it in \cref{sec:simple_algorithm}. This is followed by a description of the complete method and algorithm in \cref{sec:method}. Finally, we present results in \cref{sec:results} and discuss them in \cref{sec:discussion}.

\section{Background}\label{sec:background}

\subsection{Nested sampling}


Nested sampling \cite{nested_sampling,nested_sampling_1} is a stochastic sampling algorithm where the Bayesian evidence ($p(d|H)$ or $\evidence$) is rewritten as a one-dimensional integral in terms of the prior volume $X$
\begin{equation}\label{eq:evidence_integral}
    \evidence = \int_{0}^{1} \likelihood(X) \textrm{d}X,
\end{equation}
where $\likelihood(X)$ is the likelihood at a given prior volume $X$. If the likelihood $\likelihood(X)$ is a well-behaved function, then this formulation allows for the evidence to be approximated using an ordered sequence of decreasing prior volumes $X_i$ such that
\begin{equation}\label{eq:evidence_sum}
    \evidence \approx \hat{\evidence} = \sum_{i=1}^{N} \likelihood_{i} w_{i},
\end{equation}
where $\likelihood_i = \likelihood(X_i)$ is the likelihood at $X_i$ and the weights $w_{i}$ are, for example, given by $w_i = (1/2)(X_{i} - X_{i+1})$. The prior volume at a given iteration $X_i$ is computed in terms of the previous prior volume $X_{i-1}$, the number of points within the likelihood-constrained prior $\nlive$ and the shrinkage factor $t_i$ which is a random variable in $(0, 1)$ with probability density function $P(t) = \nlive t^{\nlive-1}$. The mean and standard deviation of $\log t$ are therefore
\begin{equation}
    \mu[\log t] = -\frac{1}{\nlive}, \qquad \sigma[\log t] = \frac{1}{\nlive}.
\end{equation}
Since each draw of $\log t_i$ is independent, the prior volume at a given iteration $i$ is approximately $X_i \approx \exp(-i/\nlive)$. We can express this as a recursive relationship in terms of $t_i$ where
\begin{equation}
    X_{i} = t_i X_{i-1}.
\end{equation}

The overall nested sampling algorithm can then be summarised as follows:
\begin{enumerate}
    \item {Draw $\nlive$ points $\{\theta_i\}_{i=1}^{\nlive} \sim \pi(\theta)$ and compute the likelihood $\mathcal{L}_i = \likelihood(\theta_i)$ of each point,}
    \item {Choose the point $\theta^{*}$ with the lowest likelihood $\mathcal{L}^{*} \equiv \mathcal{L}(\theta^{*})$,}
    \item {Draw new points $\hat{\theta}$ until $\mathcal{L}(\hat{\theta}) > \mathcal{L}^{*}$,}
    \item {Replace $\theta^{*}$ with the new point $\hat{\theta}$ and add $\theta^{*}$ to the \textit{nested samples},}
    \item {Update the evidence estimate via \cref{eq:evidence_sum},}
    \item {Repeat steps 2-5 until a stopping criterion is met.}
\end{enumerate}
The algorithm returns a set of nested samples, with corresponding prior volumes and likelihoods, and an evidence estimate with a corresponding error. The stopping criterion is typically related to the fractional change in the evidence between iterations \cite{buchner_review}.

Given a completed nested sampling run, posterior samples can be drawn by computing the posterior weights for each nested sample
\begin{equation}
    p_i = \frac{\mathcal{L}_i w_{i}}{\hat{\evidence}},
\end{equation}
and then, for example, rejection sampling can be used to obtain samples from the posterior distribution.

This formulation has been extended and modified in various works, such as to allow for a varying number of live points \cite{dynamic-ns}, to use different proposal methods \cite{polychord,nessai,multinest}, or even using different definitions of the weights $w_i$ in \cref{eq:evidence_sum} \cite{diffusive-ns,Cameron:2013,multinest_importance}, which is the focus of this work.

As mentioned previously, the main challenge when implementing a nested sampling algorithm is drawing live points that are \gls{iid} according to the prior and satisfy the likelihood constraint at the current iteration. There are various different approaches to this. In the original paper~\cite{nested_sampling_1}, Skilling proposes using \gls{mcmc} over the prior and accepting only those points for which $\likelihood(\theta) > \likelihood^{*}$ until the correlation with the starting point (one of the existing samples) has been lost. This method requires a random walk that can adapt to the continuously shrinking likelihood-constrained prior and a method for determining the number of steps to take \cite{buchner_review}. Further modifications are often needed to handle multi-modality and complex correlations between parameters, for example, as implemented in \citeauthor{lalinference}~\cite{lalinference}.
Similarly, slice sampling \cite{Neal:2003}, where samples are drawn from a randomly oriented line within the likelihood-constrained prior, has also been used~\cite{polychord}. The challenge in this case is choosing the direction of the line and how to sample from it. Another approach is to sample from a bounding (or proposal) distribution that directly approximates or contains the likelihood-constrained prior, such as ellipsoids \cite{multinest,multinest_importance} or mixtures of these to handle, for example, multi-modality. Finally, there are algorithms that use a mix of the aforementioned methods~\cite{ultranest,dynesty}.

One limitation of nested sampling is its inherently sequential nature. This is addressed in part by dynamic nested sampling \cite{dynamic-ns} where an initial exploratory run is then retroactively improved upon by adding samples in regions of interest. However, the core algorithm is still sequential. Diffusive nested sampling \cite{diffusive-ns} tackles this by using a multi-level exploration method which allows returning to lower likelihoods. We draw from this variant of nested sampling when developing our modified algorithm.

Machine learning has also been incorporated into nested sampling algorithms to address some of the limitations and accelerate inference. In \citeauthor{bambi}~\cite{bambi}, the likelihood is approximated using a neural network which, for computationally expensive likelihoods, can reduce the overall computational cost. In \citeauthor{Alsing:2021wef}~\cite{Alsing:2021wef}, normalising flows are used to allow for arbitrary priors which could otherwise not be used, for example, when using a posterior distribution as the prior for subsequent inference. Normalising flows have also been applied specifically to the proposal process. The algorithm proposed in \citeauthor{moss-ns-flows}~\cite{moss-ns-flows} improves \gls{mcmc} efficiency by transforming the sampling parameter space to a simpler space using a normalising flow and in ~\citeauthor{nessai}\cite{nessai}, we proposed \nessai which uses normalising flows to directly approximate the likelihood-constrained prior and to avoid the need for \gls{mcmc}, greatly improving sampling efficiency. We discuss \nessai in detail in \cref{sec:nessai}.

\subsection{\nessai: Nested sampling with normalising flows} \label{sec:nessai}

In ~\citeauthor{nessai}\cite{nessai}, to address the aforementioned challenged of proposing new live points from the likelihood-constrained prior, we introduced \nessai, a nested sampling algorithm that incorporates normalising flows in the proposal process. We now review the core aspects of \nessai.

Normalising flows are a family of parameterised invertible transforms that can be trained via an optimisation process to map from a simple distribution $p_{\latent}(z)$ in the latent space ($\latent$) to a complex distribution $p_{\physical}(x)$ in the data space ($\physical$). They were first proposed in \cite{Rezende:flows,Dinh:NICE} and have since been applied to a range of problems including image synthesis, noise modelling, physics and simulation-based inference \cite{Kobyzev:nf_review,Papamakarios:nf_review,Cranmer:sbi}. 

One property that distinguishes normalising flows from other generative models, such has Variational Autoencoders \cite{KingmaW13} and Generative Adversarial Networks \cite{GoodfellowPMXWO20}, is their construction allows for an explicit expression for the learnt distribution $p_{\physical}(x)$
\begin{equation}\label{eq:change_of_variable}
    p_{\physical}(x) = p_{\latent}(f(x)) \left| \textrm{det} \left( \frac{\partial f(x)}{\partial x}\right)\right|,
\end{equation}
where $f$ is the normalising flow and  $\left| \textrm{det} \left( \partial f(x) / \partial x\right)\right|$ is the Jacobian determinant. The normalising flow $f$ must be constructed such that the mapping is invertible and has a tractable Jacobian determinant. Depending on how the mapping is constructed, they fall into two main categories: \textit{autoregressive flows} and \textit{coupling flows}. The former have more expressive power at the cost of being more computational expensive to train and evaluate, whereas the opposite is true for the later \cite{Papamakarios:nf_review}. In \citeauthor{nessai}~\cite{nessai} and in this work, we use coupling flows based on RealNVP~\cite{real_nvp}. For a complete review of normalising flows, see ~\citeauthor{Kobyzev:nf_review}~\cite{Kobyzev:nf_review} and ~\citeauthor{Papamakarios:nf_review}~\cite{Papamakarios:nf_review}. 

 In \nessai, at a given iteration, a normalising flow is trained using the current live points. The trained flow maps the live points from the sampling space $\physical$ to samples in the latent space $\latent$. New samples are then drawn by sampling from a truncated latent distribution and applying the inverse mapping $f^{-1}$. Finally, rejection sampling is used to ensure that the samples are distributed according to the prior. The benefit of this approach is that all the samples are \gls{iid}, removing the need for MCMC sampling. Furthermore, since the points are drawn in parallel, the likelihood evaluation can also be parallelised, further reducing the time taken for the algorithm to converge.
 
 However, we found that the rejection sampling step can be inefficient and lead to many samples being discarded. In particular, for the results we presented in \citeauthor{nessai}~\cite{nessai}, this rejection sampling accounted for approximately \nessaipopulationtime of the total sampling time and, unlike the likelihood evaluation, this time cannot be significantly reduced via parallelisation. Additionally, we found it was necessary to reparameterise certain parameters that would otherwise be difficult to sample or make the rejection sampling inefficient. For example, parameters with posterior distributions that rail against the prior bounds could be under-sampled when the latent space is truncated. Whilst reparameterising these problematic parameters does address these issues, it requires prior knowledge of the parameter space.
 
\subsection{Alternative formulations of nested sampling}

In this section, we highlight alternative formulations of nested sampling that will be built upon in this work.

\subsubsection{Diffusive nested sampling}

Diffusive nested sampling~\cite{diffusive-ns} uses a multi-level exploration method where a mixture of constrained distributions is sampled from at each iteration using \gls{mcmc}. The constrained distributions are added sequentially and each contains approximately $e^{-1}$ of the prior volume of the previous. In contrast to standard nested sampling approaches, all the samples from the \gls{mcmc} chain are kept and those that do not meet the current likelihood criteria are added to the previous level. The values for the prior volume $X$ are estimated using the fraction of samples above the likelihood threshold compared to the total number of samples.

This variation of nested sampling avoids the strict likelihood constraint and utilises all the samples drawn at a given iteration but still requires that new points be sampled from the prior.

\subsubsection{Importance nested sampling}\label{sec:ins}

Importance nested sampling was proposed in \citeauthor{Cameron:2013}~\cite{Cameron:2013} and expanded upon in \citeauthor{multinest_importance}~\cite{multinest_importance}. In this version of nested sampling, the evidence integral is approximated in terms of a \textit{pseudo-importance sampling density} $Q(\theta)$
\begin{equation}\label{eq:is_evidence}
    \hat{\evidence} = \frac{1}{\ntotal} \sum_{i=1}^{{\ntotal}} \frac{\likelihood(\theta_i)\prior(\theta_{i})}{Q(\theta_{i})},
\end{equation}
where $\ntotal$ is the total number of nested samples. Posterior weights are then computed using
\begin{equation}\label{eq:ins_posterior}
    p_i = \frac{\likelihood(\theta_i) \prior(\theta_i)}{\ntotal Q(\theta_i)},
\end{equation}
and these can be used to obtain posterior samples via rejection sampling, or used directly in weighted histograms or kernel density estimates to approximate marginal distributions.

In standard importance sampling, the unbiased estimator for the variance of the evidence is given by
\begin{equation}\label{eq:ins_evidence_error}
    \sigma^{2}[\hat{\evidence}] = \frac{1}{\ntotal (\ntotal - 1)} \sum_{i=1}^{\ntotal} \left[ \frac{\mathcal{L}(\theta_i) \pi(\theta_i)}{Q(\theta_i)} - \hat{\evidence}\right]^{2},
\end{equation}
however, this does not apply when using a pseudo-importance sampling density, which is the case in \multinest~\cite{multinest_importance}.

In \multinest \cite{multinest,multinest_importance}, one or more ellipsoidal distributions are used to construct an approximation of the current likelihood contour defined by $\likelihood^*$. New points are then drawn from within this proposal distribution and their likelihood evaluated until $\likelihood(\hat{\theta}) > \likelihood^{*}$ and, similarly to diffusive nested sampling, all these points are used in the evidence summation and define the number of points within a level $n_i$. The pseudo-importance sampling density for each point is given by
\begin{equation}\label{eq:multinset_Q}
    Q(\theta) = \frac{1}{N_{\textrm{tot}}} \sum_{i=1}^{N_{\textrm{iter}}} \frac{n_i E_{i}(\theta)}{V_{\textrm{tot},i}},
\end{equation}
where $V_{\textrm{tot}, i}$ is the volume of the bounding distribution, $E_{i}$ is an indicator function that is 1 if the point lies within the $i$'th ellipsoidal decomposition and 0 otherwise, $N_\textrm{iter}$ is the number of iterations, where an iteration is an instance of the ellipsoidal decomposition and $N_{\textrm{tot}}$ is the total number of points $N_\textrm{tot}=\sum_{i=1}^{N_\textrm{iter}} n_{i}$.

This formulation of the evidence removes the requirement that samples are distributed according to the likelihood-constrained prior so long as the exact distribution of nested samples $Q(\theta)$ can be written down. However, only a single point is removed and updated between each update of the ellipsoidal decomposition, therefore convergence will require computing the decomposition hundreds or thousands of times. This makes it ill-suited to use with normalising flows that are, in comparison, slow to train.

\subsubsection{Nested Sampling via Sequential Monte Carlo}\label{sec:ns-smc}

\Gls{smc} is a general extension of importance sampling where random samples with corresponding weights are drawn from a sequence of probability densities such that they converge towards a target density \cite{2019arXiv190304797N}. These algorithms are typically comprised of three main steps: \textit{mutation} in which the samples are moved towards the target density via a Markov kernel, \textit{correction} where the weights of the samples are updated, and \textit{selection} where the samples are resampled according to their weights.

In \citeauthor{ns-smc}~\cite{ns-smc}, the authors draw parallels between nested sampling and \gls{smc} and show that nested sampling is a type of adaptive \gls{smc} algorithm where weights are assigned suboptimally. They also highlight several limitations of the standard nested sampling algorithm, including the assumption of independent samples. They propose a new class of \gls{smc} algorithms called \gls{ns-smc} and demonstrate that it is equivalent to nested sampling but addresses the aforementioned limitations. This formulation bares similarities to the importance nested sampling~\cite{Cameron:2013,multinest_importance} but removes batches of live points at each iteration and includes the mutation and selection steps that are typical in \gls{smc}.

A downside of this formulation is that since the points are resampled at each iteration, some samples for which the likelihood has been evaluated are discarded and not used in the final evidence estimate or output. In this work, we aim to avoid this by not including the resampling step and instead directly using the weights of the samples when constructing the next level.

\section{Core importance nested algorithm}\label{sec:simple_algorithm}

In this section, we motivate and present the core importance nested sampling algorithm used in \nessai. We extend the formulation of importance nested sampling described in \cref{sec:ins} to allow the use of normalising flows instead of ellipsoidal bounding distributions. We also draw on the design of diffusive nested sampling where the likelihood constraint is relaxed such that samples are not rejected based on their likelihood.

We start by considering the definition of the evidence from \cref{eq:is_evidence}. In importance nested sampling, the aim is to construct an importance sampling density $Q(\theta)$, which we will call \textit{meta-proposal}, from which samples can be drawn, and used to estimate the evidence. The error on this estimate is given by \cref{eq:ins_evidence_error} and depends on the number of samples $N_\textrm{tot}$ and $Q(\theta)$. If we consider a fixed number of samples, the meta-proposal that maximises the \gls{ess} of the set of summands $\likelihood(\theta_i)\pi(\theta_i) / Q(\theta_i)$, and therefore provides the most precise evidence estimate, will be $Q(\theta) \equiv \likelihood(\theta)\prior(\theta) / \evidence$, i.e. when $Q(\theta)$ is equal to the target posterior. Since the evidence is unknown a-priori, the aim is to construct the meta-proposal such that $Q(\theta) \propto \likelihood(\theta)\prior(\theta)$.

This formulation of nested sampling is closely related to Variational Inference~\cite{vi_review}, where the goal is to approximate a target probability density. In this case, the target density is $\likelihood(\theta)\prior(\theta)$ and the approximate distribution is the meta-proposal $Q(\theta)$. The difference is in how the approximate distribution is obtained. In variational inference, the approximate distribution is optimised by minimising a variational objective, whereas in this algorithm the distribution is constructed by progressively sampling and adding proposal distributions.

We now consider how to construct the meta-proposal using normalising flows. An important difference between the ellipsoidal bounds used in \multinest and normalising flows is the space over which they are defined. For a normalising flow, this depends on the domain of the latent distribution $p_{\latent}$. For the typical case of a \ndimensional{n} Gaussian the mapping is defined such that $f : \mathbb{R}^{n} \to \mathbb{R}^{n}$, so the flow will have infinite support. We need the meta-proposal to have the same support as the prior, so we include an additional invertible transform that maps from $\mathbb{R}^{n}$ to a bounded space, such as the Sigmoid $s(x) = [1 - \exp(-x)]^{-1}$. We denote the bounded space $\physical$ and the unbounded space $\physical'$.

Therefore, instead of considering a series of bounded distributions, we consider a set of $N$ normalised proposal distributions (normalising flows) $\{q_1, ..., q_{N}\}$ all defined over the entire prior volume and with corresponding weights $\alpha_j$ defined such that $\sum_{j=1}^{N} \alpha_j =1$.
The overall proposal density as a function of $\theta$ is given by
%
\begin{equation}\label{eq:meta_proposal}
    Q(\theta) = \sum_{j=1}^{N} \alpha_{j} q_{j}(\theta).
\end{equation}
In practice, in order to sample from $Q(\theta)$ we first draw a proposal $k\in\{1,\ldots,N\}$, drawn from a categorical distribution with category weights $\{\alpha_1,...,\alpha_N\}$, then a sample is drawn from the sub-proposal $q_k(\theta)$.

With this formulation, we can compute an estimate of the evidence for a set of samples drawn from $Q(\theta)$ using \cref{eq:is_evidence} and, as noted in \citeauthor{multinest_importance}~\cite{multinest_importance}, we no longer require new samples that have monotonically increasing likelihood values. Furthermore, as described in \citeauthor{ns-smc}~\cite{ns-smc}, we do not require that new samples be \gls{iid} or distributed according to the likelihood-constrained prior. This removes the need for the rejection sampling that was a bottleneck in the version of \nessai we described in \citeauthor{nessai}~\cite{nessai}.

We now outline a simplified importance nested sampling algorithm which we build upon in later sections. The main changes are to steps 2-5 of the standard nested sampling algorithm outlined in \cref{sec:background}. Instead of removing a point and finding a single replacement point, we construct a proposal distribution $q_j(\theta)$ based on the points sampled thus far and draw a set of $\nperflow$ new points $\Theta_j = \{\theta_i\}_{i=1}^{\nperflow}$ which are added to the overall set of points $\{\Theta_1, ..., \Theta_{j-1}\}$. The meta-proposal $Q(\theta)$ is then updated to include $q_j(\theta)$ and the evidence is updated. The new importance nested sampling algorithm therefore consists of the following steps:

\begin{enumerate}
    \item {Draw $\nlive$ points $\{\theta_i\}_{i=1}^{\nlive} \sim \pi(\theta)$ and compute the likelihood $\mathcal{L}_i = \likelihood(\theta_i)$ of each point, }
    \item {add the next proposal distribution $q_j(\theta)$,}\label{step:add_proposal}
    \item {draw $\nperflow$ samples from $\Theta_j = \{\theta_i\}_{i=1}^{\nperflow} \sim q_j(\theta)$ and compute the corresponding likelihoods, }\label{step:sample_proposal}
    \item {update the meta-proposal $Q(\theta)$ to include $q_j(\theta)$}\label{step:update_meta_proposal},
    \item {compute the evidence and the corresponding error via \cref{eq:is_evidence,eq:ins_evidence_error},}\label{step:update_evidence}
    \item {repeat steps 2-5 until a stopping criterion is met,}
    \item{redraw independent samples from the final meta-proposal,}
    \item {compute the final evidence and posterior weights using the independent samples and \cref{eq:is_evidence,eq:ins_posterior}.}
\end{enumerate}
This includes an additional step not present in standard nested sampling: redrawing independent samples from the final meta-proposal. Since subsequent proposals are constructed using samples from the previous iterations, new samples are not \gls{iid} and \cref{eq:is_evidence,eq:ins_evidence_error,eq:ins_posterior} do not strictly apply. However, once the meta-proposal is finalised, \gls{iid} samples can be sampled and used to compute unbiased estimates of the evidence and posterior weights.

The design of the algorithm hinges on how the next proposal distribution is added, how the number of samples drawn from each proposal ($\nperflow$) is determined and how the weights in the meta-proposal $Q(\theta)$ are determined. Note that the first proposal distribution $q_0(\theta)$ will typically be the prior.
We now apply this simplified algorithm to a toy example.

\subsection{Toy example}\label{sec:toy_example}

\begin{figure*}
    \begin{indented}\item[]
        \centering
        \includegraphics[width=\figwidth]{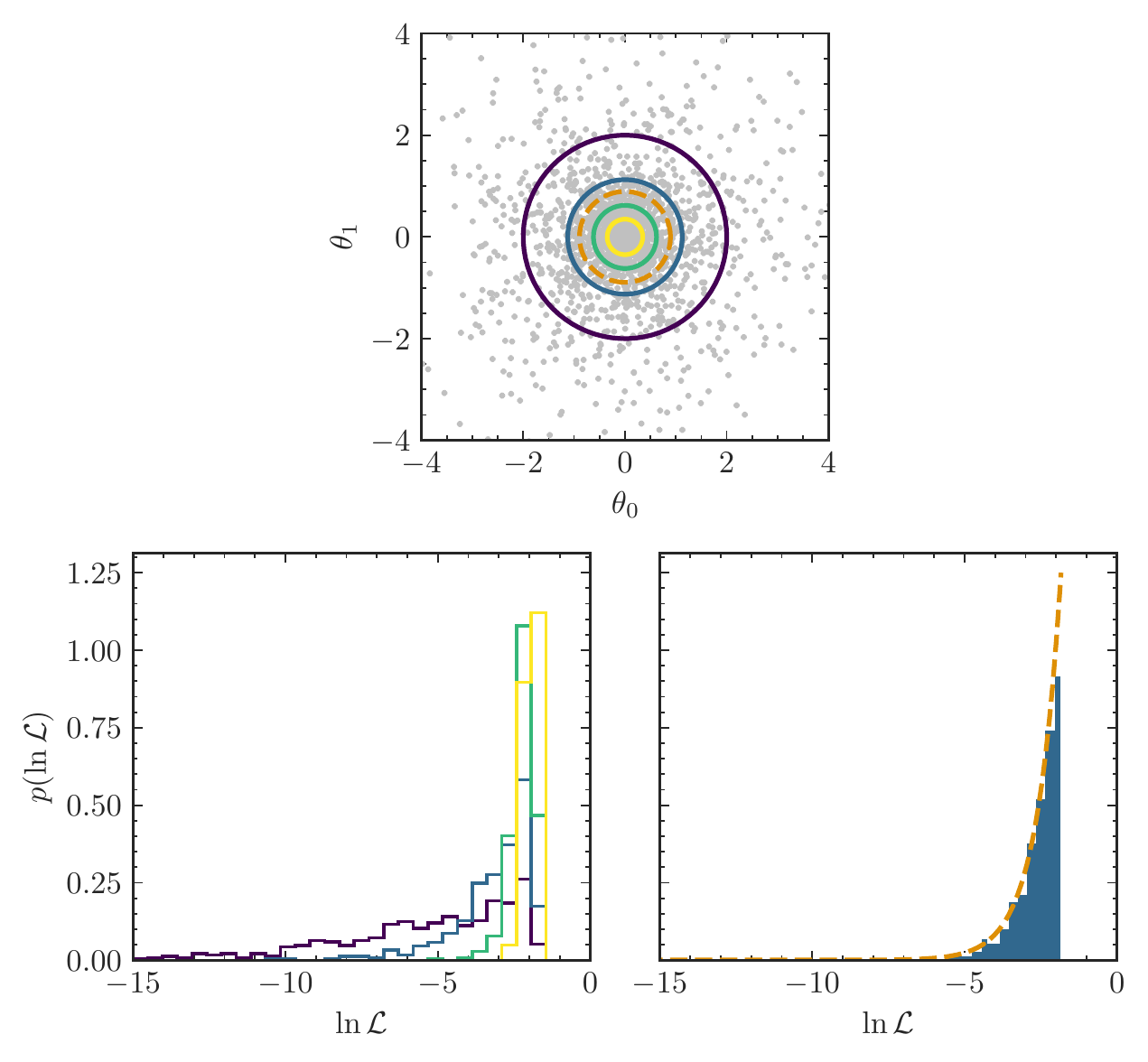}
    \end{indented}
    \caption{Results for the toy example described in \cref{sec:toy_example}. \textbf{Top:} the final samples are shown in grey, the solid lines show the 1-$\sigma$ contour for each proposal distribution starting with the prior, lighter colours indicate later iterations. The orange dashed line shows the  1-$\sigma$ contour for the analytic posterior distribution. \textbf{Bottom left:} distribution of log-likelihoods for the final samples drawn from each proposal distribution. \textbf{Bottom right:} distribution of the log-likelihoods of the final samples weighted by their corresponding posterior weights. The orange dashed line indicates the analytic posterior distribution computed using \cref{eq:post_likelihood}.}
    \label{fig:toy_example}
\end{figure*}

In this toy example, we consider a simple problem with an analytic evidence and posterior distribution. We apply the algorithm described in \cref{sec:simple_algorithm} but with some simplifications. This allows us to validate the core algorithm.

We use a \ndimensional{\toyexamplendims} Gaussian likelihood with mean $\mu_{\likelihood}=0$ and standard deviation $\sigma_{\likelihood}=1$ and a Gaussian prior a with mean $\mu_{\prior}=0$ and standard deviation $\sigma_{\prior}=\toyexamplepriorstd$. The posterior distribution is therefore another Gaussian distribution with mean $\mu_{Post}=0$ and standard deviation $\sigma_{Post} = \sqrt{1 / [(1 / \sigma_{\likelihood}^{2}) + (1 / \sigma_{\pi}^2)]}$. The evidence is given by a Gaussian distribution with mean $\mu_{\prior}$ and standard deviation $\sqrt{\sigma_{\likelihood}^2 + \sigma_{\prior}^2}$ evaluated at $\mu_{\likelihood}$, so $\evidence_\textrm{Analytic}=\toyexampleanalyticevidence$.

To make the comparison between the true and sampled posterior distributions easier, we express the posterior distribution in terms of the log-likelihood $p(\ln \likelihood)$. To do this, we note that the posterior distribution defined in terms of the radius squared is $p(r^2) = \chi^2_2 (r^2) / \sigma_{Post}^2$ where $\chi^2_2$ is a chi-squared distribution with two degrees of freedom. Then
\begin{equation}\label{eq:post_likelihood}
    p(\ln\likelihood) = p(r^2) \left| \frac{\partial r^2}{\partial \ln \likelihood}\right| ,
\end{equation}
where
\begin{equation}
    r^2 = -2 \sigma_{\likelihood}^2 \left[\ln(2\pi\sigma_{\likelihood}^2) + \ln \likelihood \right],
\end{equation}
which is defined on $[0, \infty)$ since the maximum possible value of the log-likelihood is $\ln\likelihood = -\ln(2\pi\sigma_{\likelihood}^2)$.

The four steps we must define for the simplified algorithm are: how to construct each proposal distribution, how to determine the number of samples to draw from each proposal, how to determine the weights for each proposal in the meta-proposal and a stopping criterion. For the proposals, instead of normalising flows, we use \ndimensional{2} Gaussian distributions $q_j(\theta)$ with mean zero and different standard deviations. We determine the standard deviation of each proposal by setting a likelihood threshold $\likelihoodthreshold$ such that 50\% of the points from the previous iteration are discarded and then compute the standard deviation of the remaining points. We set the number of samples drawn from each proposal to constant $\nperflow = \nlive = \checked{500}$ and set the weights for the meta-proposal $\alpha_j$ to be equal. This means that each proposal will contribute equally to the meta-proposal. Finally, instead of using a stopping criterion, we define a fixed number of proposal distributions (iterations) $N=\toyexamplendist$ where the first is the prior distribution $q_0(\theta) \equiv \prior(\theta)$. This is akin to fixing the number of iterations in a normal nested sampling algorithm. Once the final proposal has been added, we draw \gls{iid} samples from the finalised meta-proposal and compute the final unbiased evidence estimate and posterior weights.

We present the results obtained with this algorithm in \cref{fig:toy_example}. This shows the samples and the 1-$\sigma$ contours for each of the proposal distributions, along with the corresponding distribution of log-likelihoods. We compute two evidence estimates: one with the initial samples that are not \gls{iid} $\hat{\evidence} = \toyexampleevidence \pm \toyexampleevidenceerror$ and the other with the final \gls{iid} samples $\hat{\evidence} = \toyexampleevidencefinal \pm \toyexampleevidenceerrorfinal$. We find that both are in agreement with the analytic value, $Z = \toyexampleanalyticevidence$, but, as we will see in \cref{sec:analytic_likelihoods}, the initial estimate will be biased, the bias is just very small in this simple example.
This demonstrates that the underlying algorithm can reliably estimate the evidence. We also compute the posterior weights using \cref{eq:ins_posterior} and plot the weighted histogram in log-likelihood space, which shows good agreement with the analytic expression from \cref{eq:post_likelihood}. Overall, these results demonstrate the principles of the proposed algorithm and that, for a simple toy example, it converges to the expected result.

\section{Method}\label{sec:method}

Having outlined the underlying algorithm, we now describe each of the steps in the complete algorithm in detail.

\subsection{Constructing proposal distributions}\label{sec:level_method}

With this formulation of nested sampling, the main design choice is how to construct the proposal distribution $q_{j}(\theta)$ at each iteration (step 2). This is akin to drawing new samples in standard nested sampling however, since we no longer require an ordered sequence of points with decreasing prior volume, new points no longer need strictly increasing likelihood values.

The new proposal $q_j(\theta)$ at each iteration is defined in terms of a likelihood threshold $\likelihoodthreshold$: of the current $\nlive$ points, $\nremoved$ are discarded based on a likelihood threshold and the remaining $\nlive - \nremoved$ points are used to construct the next proposal distribution $q_j(\theta)$. In our implementation, this is done by training a normalising flow. 
The result is a series of increasingly dense proposal distributions, which is equivalent to the distributions becoming narrower in the log-likelihood space. This is shown in \cref{fig:toy_example}.

We therefore require a method for determining the likelihood threshold $\likelihoodthreshold$ used to determine how many points will be discarded before constructing the next proposal distribution. We consider two methods, both of which use weights
\begin{equation}\label{eq:level_weights}
    w_{i} = \frac{\pi(\theta_i)}{Q(\theta_i)},
\end{equation}
which quantify the relative importance of each sample $\theta_j$ compared to the prior. Additionally, one could include the likelihood in the weights, however, we leave this for future work.

In the first method, the threshold $\likelihoodthreshold$ is determined using the ($1-\rho$) quantile of the likelihood values of the samples from the previous iteration, where $\rho$ is set by the user. To account for non-prior distributed samples used in our algorithm, we use a weighted quantile, where the weights are given by \cref{eq:level_weights}. This method is based on the standard method used in \gls{smc}~\cite{ns-smc} and diffusive nested sampling~\cite{diffusive-ns}, but with the addition of the weighted quantile.

The second method we consider is closely related to the first but uses log-weights $\log w_i$ instead of $w_i$. We consider the normalised sum of $\log w_i$ for the set of $N$ live points ordered by increasing likelihood
\begin{equation}
    \lambda(M) = \frac{\sum_{m=1}^{M}\log w_m}{\sum_{i=1}^{N} \log w_i},
\end{equation}
where $M$ is the number of live points to be discarded.
We then determine the value of $M$ at which $\lambda(M) \geq \rho$, for $\rho \in [0, 1]$ and set $\likelihoodthreshold \equiv \likelihood(\theta_M)$. This is analogous to shrinking the log-prior volume by a factor $\rho$ at each iteration whilst also accounting for the different weights of each sample. In practice, since the normalising flows have support over the entire prior volume, this results in increasing the entropy of $q_j(\theta)$. We therefore denote this as the \textit{entropy-based} method to distinguish it from the quantile-based method.

For both methods, we employ a maximum number of live points that can be removed - this prevents the remaining live points being too few to robustly train the next normalising flow. This maximum together with the value of $\rho$ will determine the total number of samples used in the algorithm. We also employ a minimum number of samples to ensure a minimum change in distribution of training data between subsequent proposals. We discuss the advantages and disadvantages of both methods in \cref{app:level}.

\subsection{Training normalising flows with weights}

As discussed in \cref{sec:ns-smc}, it is common practice in \gls{smc} to resample at each iteration prior to the mutation step. Different sampling methods can be used, but they all keep the total number of samples constant by including repeated samples.
This works when the mutation step is a Markov kernel, but in this work we use a normalising flow to perform the equivalent of the mutation step and, when training a normalising flow duplicates in the training data, can be problematic. In extreme cases, where only a few samples are representative, the training data could contain tens of copies of the same sample, which will make training unstable.

Without a step that is equivalent to resampling, deficiencies in training can have a cumulative effect. For example, if the mapping learnt by the normalising flow $q_j(\theta)$ under-samples a region of the space compared to the target, then if another normalising flow $q_{j+1}(\theta)$ is trained with samples drawn using $q_j(\theta)$ then $q_{j+1}(\theta)$ will also under-sample the same region. To counteract this effect, we include weights in the approximation of \gls{kld} used to train the normalising flow. We describe this in detail in \cref{app:weighted_kl}. To train the $j$-th flow, we use all samples from the current meta-proposal $Q_{j-1}(\theta)$ that satisfy the likelihood constraint $\likelihood(\theta) > \likelihoodthreshold$ and then minimise 
\begin{equation}\label{eq:weighted_kl}
    \textrm{Loss} = -\frac{1}{N} \sum_{i=1}^{N} w_i\log q_{j}(\theta_i),
\end{equation}
where $q_{j}(\theta)$ is given by \cref{eq:change_of_variable} and $w_i$ are the weights for each sample. In principle these weights could include the likelihood, however in this work we use the weights given by \cref{eq:level_weights} which are proportional to the ratio of the likelihood-constrained prior and the likelihood-constrained meta-proposal.


\subsection{Drawing samples from the proposal distributions}

At a given iteration $j$, once the normalising flow $q_j(\theta)$ has been trained (step 2), we sample from the flow (step 3) and evaluate the likelihood for each new sample. This involves sampling from the latent distribution $p_{\latent}(z)$ and then applying the inverse flow mapping $f^{-1}$ to obtain samples in $\physical'$. These samples must then be mapped backed to the original space $\physical$, where the likelihood can be computed.

The number of samples drawn at a given iteration $\nperflow$ should be determined by drawing from a multinomial distribution with $N$ possible outcomes (the number of proposal distributions) and $\ntotal = \sum_{j=1}^{N}\nperflow$ trials, however the weights for each outcome are not known prior to sampling.
Instead, we set $\nperflow$ and determine the weight for the current iteration $\alpha_j$ based on its value. We allow $\nperflow$ to either be equal to the number of samples removed at that iteration ($\nremoved$) or kept constant ($\nperflow = \nlive$). The former will maintain a fixed number of live points $\nlive$ throughout the run whereas the latter allows for $\nlive$ to vary. We discuss the consequences of this approximation in \cref{sec:updating_meta_propoal,sec:redrawing_samples}. 

Similarly to diffusive nested sampling, all the samples are kept irrespective of their likelihood,
which means that samples can ``leak'' below the current likelihood threshold. 

\subsection{Updating the meta-proposal}\label{sec:updating_meta_propoal}

Having drawn samples from the current proposal distribution, the meta-proposal $Q(\theta)$ must be updated. The overall form of $Q(\theta)$ will depend on the weights $\alpha_j$ that are assigned to each proposal. Whilst adding proposals, we approximate the weights as $\alpha_{j} \propto \nperflow$ and normalise them such that they sum to one. This approximation can be corrected for once the sampling has been terminated by fixing the weights to their values from sampling, recomputing $\nperflow$ by sampling from a multinomial distribution with weights $\{\alpha_0,...,\alpha_{\nperflow}\}$ and drawing new samples from each $q_j(\theta)$ according to $\nperflow$. However, in practice, we find error introduced by this approximation to be significantly smaller than the overall error of the estimated evidence.

\subsection{Stopping criterion}\label{sec:stopping_criterion}

We define the stopping criterion to be the ratio of the evidence between the live points and the current evidence
\begin{equation}\label{eq:stopping}
    \textrm{Condition} = \frac{\hat{Z}_{\textrm{LP}}}{\hat{Z}},
\end{equation}
where $\hat{Z}_{\textrm{LP}}$ is computed using \cref{eq:is_evidence} and including only the live points in the sum. The algorithm will then terminate when the condition is less than a user-defined threshold~$\tau$.

This is more suitable than the fractional change in the evidence between iterations, that is used in standard nested sampling algorithms, because multiple points are removed simultaneously at each iteration, the number of points can vary between iterations and points can leak below the current $\likelihoodthreshold$, which all mean fractional change does not decrease smoothly and instead can fluctuate significantly between iterations.

\subsection{Posterior samples}

Similarly to \gls{smc} and \multinest, our algorithm returns samples $\{\theta_i\}_{i=1}^{\ntotal}$ and their corresponding posterior weights $p_i$ given by \cref{eq:ins_posterior}. Different methods can then be employed to draw posterior samples. The standard approach in nested sampling is to use rejection sampling \cite{nessai} or multinomial resampling \cite{dynesty} to resample the nested samples using the posterior weights. Alternatively, the weights can be used directly in weighted histograms or kernel density estimates.

When using multinomial resampling or the weights directly, the posterior samples are not statistically independent, so it is informative to compute Kish's \gls{ess}~\cite{kish_ess}
\begin{equation}\label{eq:ess}
    \textrm{ESS} = \frac{\left[\sum_{i=1}^{N} p_i\right]^{2}}{\sum_{i=1}^{N} p_i^{2}},
\end{equation}
where $p_i$ is given by \cref{eq:ins_posterior}. This gives an indication of the effective number of posterior samples in the posterior and allows for comparing results obtained via different sampling methods. It can also be used to diagnose poorly converged runs, since a low \gls{ess} is an indication that the samples and their corresponding weights are a poor match for the true posterior distribution.

\subsection{Post-processing}\label{sec:redrawing_samples}

Once sampling is complete, we correct for the approximation of the meta-proposal $Q(\theta)$ discussed in \cref{sec:updating_meta_propoal} by redrawing $N_\textrm{Final}$ samples from the meta-proposal according the draws from the multinomial distribution. The number of samples can be equal to $N_\textrm{Total}$ or can be increased or decreased depending on the desired output.

This has the additional benefit of allowing more samples to be drawn after sampling has completed and can be used to obtain more posterior samples or decrease the estimated error on the evidence.

\subsection{Complete algorithm}\label{sec:algorithm}

We can now combine all these elements into a complete algorithm which is shown in \cref{alg:inessai}. The algorithm incorporates normalising flows but no longer requires that samples drawn from them be \gls{iid} according to the prior. Furthermore, samples are drawn and their likelihoods evaluated in batches and all the samples are kept irrespective of their likelihood. Finally, the evidence is a simple sum, so it can be updated for batches of samples. Thus, this algorithm meets all the criteria that were initially set out.

\begin{algorithm}
\caption{Overview of \inessai}\label{alg:inessai}
\KwInput{Likelihood $\mathcal{L}$, Prior $\pi$, Tolerance $\tau$, Method for determining $\nperflow$, $N_\textrm{Final}$}
\KwOutput{Evidence $\hat{Z}$, samples $\{\Theta_1, ..., \Theta_j\}$ and posterior weights $W$}
$j \gets 1$ \;
$\Theta_{1} \gets \{\theta_{i} \sim \pi\}_{i=1}^{N_1}$\;
$\ntotal \gets N_1, q_{1} \gets \pi$ \;
\While{$condition \geq \tau$}{
    $j \gets j + 1$ \;
    $q_j \gets$ trained normalising flow\;
    $\nperflow \gets$ determined via specified method\;
    $\Theta_{j} \gets \{\theta_i \sim q_j\}_{i=1}^{\nperflow}$\;
    $\ntotal \gets \ntotal + \nperflow$\;
    $\hat{Z} \gets \frac{1}{\ntotal}\sum_{i=1}^{N_{tot}}\frac{\mathcal{L}(\theta_i)\pi(\theta_i)}{Q(\theta_i)}$\;
    $W \gets \left\{\frac{\mathcal{L}(\theta_i) \pi(\theta_i)}{\ntotal Q(\theta_i)}\right\}_{i=1}^{N_\textrm{tot}}$\;
}
Redraw $N_\textrm{Final}$ samples from the final meta-proposal and compute the final evidence estimate and posterior weights.
\end{algorithm}


\subsection{Biases}\label{sec:biases}


In our algorithm, the proposal distributions (normalising flows) are trained and then sampled from, rather than being constructed post sampling. This means that, unlike in \multinest, the meta-proposal distribution is an importance sampling density and \cref{eq:ins_evidence_error} should give a reliable estimate of the evidence error. We verify this in \cref{sec:analytic_likelihoods}. 

We also note that a different bias in the evidence arises from evaluating each normalising flow with samples that were also used to train it. This is necessary since the meta-proposal requires evaluating each normalising flow on every sample. This is a side effect of the small amount of training data available to each flow and difficulty in setting the hyperparameters for $N$ different normalising flows prior to sampling. This bias is corrected for when the samples are redrawn as described in \cref{sec:redrawing_samples} which we demonstrate in \cref{sec:results}.

\section{Related work}

As described in \cref{sec:background}, the proposed method draws from existing variations of nested sampling: the soft likelihood constraint from diffusive nested sampling \cite{diffusive-ns}, the formulation of importance nested sampling used in \multinest \cite{multinest_importance} and the use of normalising flows as described in \citeauthor{nessai}~\cite{nessai} and ~\citeauthor{moss-ns-flows}~\cite{moss-ns-flows}. However, it also has parallels to standard importance sampling and the methods derived from it.

Considering the use of a sequence of normalising flows to approximate a target (or posterior) distribution, the most closely related works are Nested Variational Inference~\cite{nested_vi}, Annealed Flow Transport Monte Carlo~\cite{annealed_flow_transport_mc} and Preconditioned Monte Carlo \cite{Karamanis:PMC}. The first is a hybrid between Variational Inference and \gls{smc} where a series of parameterised distributions are simultaneously optimised using an annealed version of the target distribution. In the latter two works, the standard \gls{smc} algorithm is modified to include an additional step that uses a normalising flow. Additionally, in \citeauthor{Karamanis:PMC}~\cite{Karamanis:PMC} the authors apply their algorithm to gravitational-wave inference, however only a single simulated event is analysed rather than a set of events.

As with any stochastic sampling algorithm for Bayesian inference, this work can also be compared to simulation-based or likelihood-free inference~\cite{Cranmer:sbi} where the posterior distribution is approximated using repeated simulations of the data instead of evaluating the likelihood. This technique has been applied to data analysis in physics and astrophysics, including but not limited to gravitational-wave data analysis \cite{Gabbard:2019rde,Chua:2019wwt,Green:2020hst,Dax:2021tsq}, cosmology~\cite{2019MNRAS.488.4440A,2021MNRAS.501..954J} and particle physics~\cite{2021NatRP...3..305B}. The approach used in these methods involves training on a dataset that is representative of the entire parameter space and then being able to perform inference for any given point in that space. This is the opposite to the approach employed in this work, where the algorithm is general purpose and is not trained for a specific task but instead is trained on the fly, removing the need for expensive initial training at the cost of being slower when performing inference.

\section{Results} \label{sec:results}

We present results obtained using the algorithm described in \cref{sec:algorithm} on range of problems. We implement the algorithm in the \nessai software package and it is available at \cite{nessai_doi}. To distinguish it from the version of \nessai described in \citeauthor{nessai}~\cite{nessai}, we will refer to it as \inessai.

We run all our experiments using normalising flows based on RealNVP \cite{real_nvp} as we find that more complex flows, such as Neural Spline Flows \cite{nsf}, over-fit to the small amount of data available\footnote{A single instance of over-fitting across all the flows will not significantly impact results, however, if the flows consistently over-fit then the final result will be over-constrained.} and, compared to the other components of the algorithm, are too computationally expensive to justify using.  Furthermore, \inessai requires storing the normalising flow for each level so using a flow with more parameters can significantly increase the memory footprint of the algorithm. 

We start with a series of tests using analytic likelihoods followed by a test using a more challenging likelihood and compare these results to those obtained with \nessai. We then apply \inessai to two different gravitational-wave analyses. Finally, we investigate parallelisation of the algorithm and how it scales with the number of live points.

For all experiments, we use the entropy-based method for constructing each proposal distribution described in \cref{sec:level_method} with $\rho=0.5$. We discuss this choice in \cref{app:level}. We also set the number of samples per flow to a constant $\nperflow=\nlive$. Code to reproduce all the experiments is available at \url{https://doi.org/10.5281/zenodo.8124198}~\cite{paper-repo-doi}.

\subsection{Validation using analytic likelihoods} \label{sec:analytic_likelihoods}

We start by validating \inessai using likelihoods for which the evidence can be computed analytically in $n$ dimensions. We choose to analyse the simple case of an \ndimensional{n} Gaussian. For a more complex case, we employ the \ndimensional{n} $M$-component Gaussian mixture likelihood described and used in \citeauthor{moss-ns-flows}\cite{moss-ns-flows} and \citeauthor{dynamic-ns}\cite{dynamic-ns}
\begin{equation}
    \likelihood_{\textrm{GM}}(\theta) = \sum_{m=1}^{M} W^{(m)} \left(2\pi {\sigma^{(m)}}^{2} \right)^{-n/2} \exp \left( \frac{-|\theta - \mu^{(m)}|^2}{2{\sigma^{(m)}}^{2}} \right),
\end{equation}
where $\mu^{(m)}$ and $\sigma^{(m)}$ are the mean and standard deviation of each component in all dimensions and $\sum_{m=1}^{M} W^{(m)} = 1$. We use the same hyperparameters~\cite{moss-ns-flows,dynamic-ns}: $M=4$, $W^{(m)} = \{0.4, 0.3, 0.2, 0.1 \}$, $\mu_{1}^{(m)} = \{0, 0, 4, -4\}$, $\mu_{2}^{(m)} = \{4, -4, 0, 0\}$,  $\mu_{n}^{(m)} = 0 \;\forall\; n \in \{3, ..., n\}$ and $m \in \{1,...,M\}$, and $\sigma^{(m)}=1\;\forall\;m \in {1,...,M}$.  

\begin{figure}
    \begin{indented}\item[]
        \centering
        \includegraphics{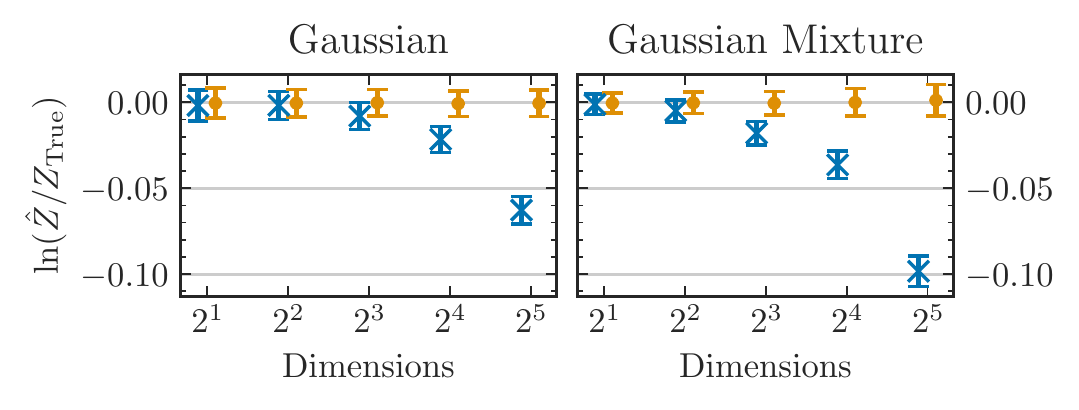}
    \end{indented}
    \caption{Mean estimated log-evidence before (blue cross) and after (orange dot) the resampling step described in \cref{sec:redrawing_samples} for an \protect\ndimensional{n} Gaussian and Gaussian Mixture. The error-bars show the mean estimated error for the log-evidence. The estimated evidence has been rescaled using the true value such that the distributions of log-evidences should be centred around zero. The number of samples drawn during the resampling step is set such that is equal to the number of samples accumulated during the initial sampling.}
    \label{fig:evidence_correction}
\end{figure}

For both likelihoods, we consider $n=\{ 2, 4, 8, 16, 32\}$ and use uniform priors on $[-10, 10]^{n}$. The analytical log-evidence for both models is $\ln Z = - n \log 20$. We analyse each likelihood \nanalytictests times, including redrawing the samples as described in \cref{sec:redrawing_samples}, and examine the distribution of the log-evidence estimates and the corresponding estimated error. In \cref{fig:evidence_correction}, we include the result of the redrawing of the samples and recomputing the final log-evidence estimate. This shows that without redrawing the samples there is a bias in the estimated log-evidence, however this bias is small compared to the value of the log-evidence, for example, for the \ndimensional{32} Gaussian and Gaussian Mixture the true log-evidence is -95.86 and the average biases are \averagebias and \averagebiasgmm respectively. After redrawing the samples, \inessai reliably estimates the evidence for both models for all values of $n$. We also compare the distribution of the re-computed log-evidences alongside the expected distribution computed using \cref{eq:ins_evidence_error} in \cref{app:validating_variance} and observe that the estimated log-evidence errors agree with the observed distributions.

\subsection{Comparison with standard nested sampling}\label{sec:comparison_to_nessai}

\begin{figure}
    \begin{indented}\item[]
        \centering
        \includegraphics[width=\figwidth]{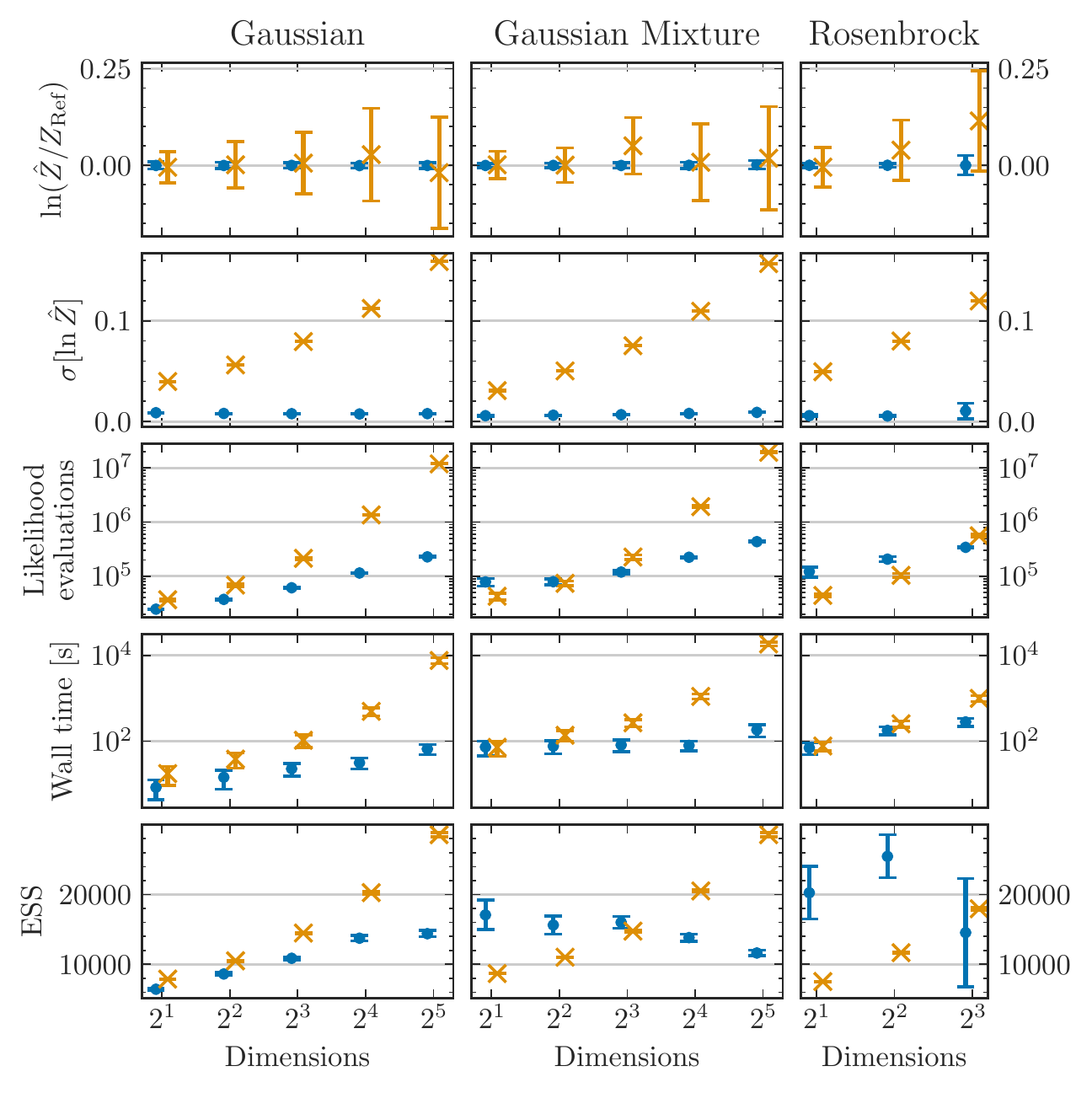}
    \end{indented}
    \caption{Comparison of results produced using \nessai (orange) and \inessai (blue) when applied to the \protect\ndimensional{n} Gaussian, Gaussian Mixture and  Rosenbrock likelihoods as described in \cref{sec:analytic_likelihoods,sec:rosenbrock}. From top to bottom, results are shown for the final estimated log-evidence rescaled by a reference evidence (the true value for the Gaussian and Gaussian Mixture and the mean value obtained with \inessai for the Rosenbrock), the estimated log-evidence error, the total number of likelihood evaluations, the total wall time in seconds and the \glsfirst{ess} of the posterior distribution. Results are averaged over \nanalytictests runs with different random seeds for both samplers and the error bars show the standard deviation.}
    \label{fig:comparison}
\end{figure}

We now compare \inessai with standard nested sampling, in particular the standard version of \nessai. This allows us to verify the results obtained with \inessai, compare the observed and estimated evidences and evidence errors, the number of likelihood evaluations, the wall time and \gls{ess} of the posterior distribution. We repeat the analyses described in \cref{sec:analytic_likelihoods} using \nessai and present the results for both likelihoods in \cref{fig:comparison}.

\Cref{fig:comparison} shows that \inessai produces estimates of the log-evidence for the Gaussian and Gaussian Mixture that are consistent with \nessai but have significantly lower variances and the corresponding estimates of the error are correspondingly smaller. We explore how the error on the log-evidence estimate scales in \cref{sec:scaling}. Furthermore, \cref{fig:comparison} shows that \inessai requires a comparable number of likelihood evaluations in lower dimensions but more than an order of magnitude less in higher dimensions and a similar trend is seen with the wall time. However, this behaviour is highly dependent on the user-defined settings, which in these experiments were set based on the requirements for the high-dimensional analyses. The \gls{ess} of the posterior distribution highlights a notable difference between the two samplers; with \nessai the \gls{ess} increases as the number of dimensions increase for both likelihoods whereas with \inessai, for the Gaussian Mixture likelihood, it decreases in higher dimensions but is still of order $10^4$. Since in importance nested sampling the \gls{ess} depends on how well the meta-proposal approximates the likelihood times the prior, a lower \gls{ess} indicates a ``worse'' approximation. In contrast, in standard nested sampling, and therefore \nessai, the \gls{ess} does not depend on the convergence of the sampler and an under- or over-constrained result can still have a large \gls{ess}.

\subsection{Testing on more challenging likelihoods}\label{sec:rosenbrock}

To further test \inessai, we consider the \ndimensional{n} Rosenbrock likelihood~\cite{rosenbrock} which has highly correlated parameters and is recognised as a challenging function to sample. We use the more involved variant~\cite{Goldberg1988,shang_rosenbrock} where the log-likelihood is defined as 
\begin{equation}\label{eq:rosenbrock}
    \ln \mathcal{L}_{\textrm{Rosenbrock}} (\theta) = - \sum_{i=1}^{n-1} [100 (\theta_{i+1} - \theta_i^2) ^ 2 + (1 - \theta_{i})^{2}],
\end{equation}
with a uniform prior on $[-5, 5]^{n}$. We test for $n=\{2, 4, 8\}$ and run \inessai~\checked{50} times for each $n$. Above $n=2$ there is no analytical solution for the log-evidence of the Rosenbrock likelihood, so we compare results to those obtained with \nessai. We present these results in \cref{fig:comparison}. We observe that \inessai is consistent with \nessai for $n=2$ but for $n=\{4, 8\}$ predicts a lower evidence than \nessai, however the relative difference is less than 1\%. The number of likelihood evaluations and wall times are comparable between both samplers but \inessai has a larger \gls{ess} in $n=\{2, 4\}$ and lower in $n=8$. To better understand these differences, we inspect the results obtained with \nessai and find that the insertion indices \cite{Fowlie:2020mzs,nessai} are consistent with the results being over-constrained (see \cref{app:rosenbrock_p_values}). This corresponds to the log-evidence being marginally over-estimated which agrees with the differences in estimated log-evidence observed in \cref{fig:comparison}.

\subsection{Probability-probability test with binary black hole signals}\label{sec:bbh_results}

\begin{figure}
    \begin{indented}\item[]
        \centering
        \includegraphics[width=\linewidth]{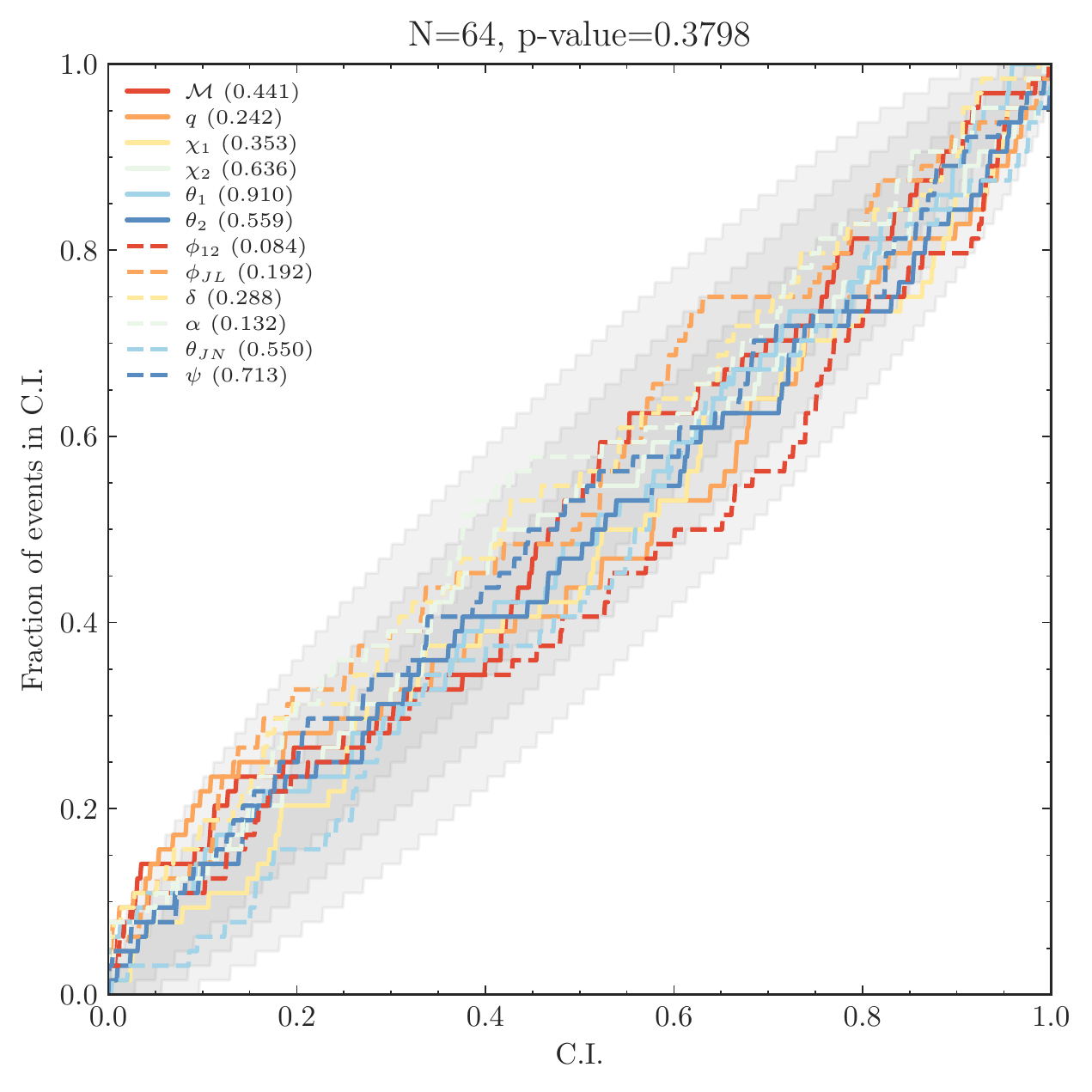}
    \end{indented}
    \caption{Probability-probability plot for \numbbhsignals simulated signals from binary black hole coalescence analysed using \inessai. The shaded regions indicated the 1-, 2- and 3-$\sigma$ confidence intervals. Individual $p$-values are shown for each parameter and the combined $p$-value is also shown.}
    \label{fig:bbh_pp_plot}
\end{figure}

As a more practical test for \inessai, we repeat the analysis used to validate \nessai in \citeauthor{nessai}~\cite{nessai}, where we used \bilby\cite{bilby} and \nessai to analyse simulated signals from compact binary coalescence of binary black holes injected into \bbhduration seconds of data sampled at \qty{\bbhsamplingfrequency}{Hz} in a three-detector network. For this analysis, we use the same priors (described in Appendix C of~\citeauthor{nessai}~\cite{nessai}) and enable phase, distance and time marginalisation in the likelihood. This reduces the parameter space to 12 parameters. We analyse \numbbhsignals injections simulated from the same priors and produce a probability-probability (P-P) plot and corresponding $p$-values using \bilby. This analysis includes the resampling step described in \cref{sec:redrawing_samples} and we re-draw the same number of samples that were used in the initial sampling, doubling the number of likelihood evaluations. The probability-probability plot is presented in \cref{fig:bbh_pp_plot} with individual and combined $p$-values. The combined $p$-value is \combinedpvalue which demonstrates that \inessai reliably recovers all 12 parameters. Furthermore, these results are obtained without introducing any of reparameterisations used in standard \nessai~\cite{nessai} to handle, for example, angles and spin magnitudes.

\begin{figure}
    \begin{indented}\item[]
        \centering
        \includegraphics[width=\figwidth]{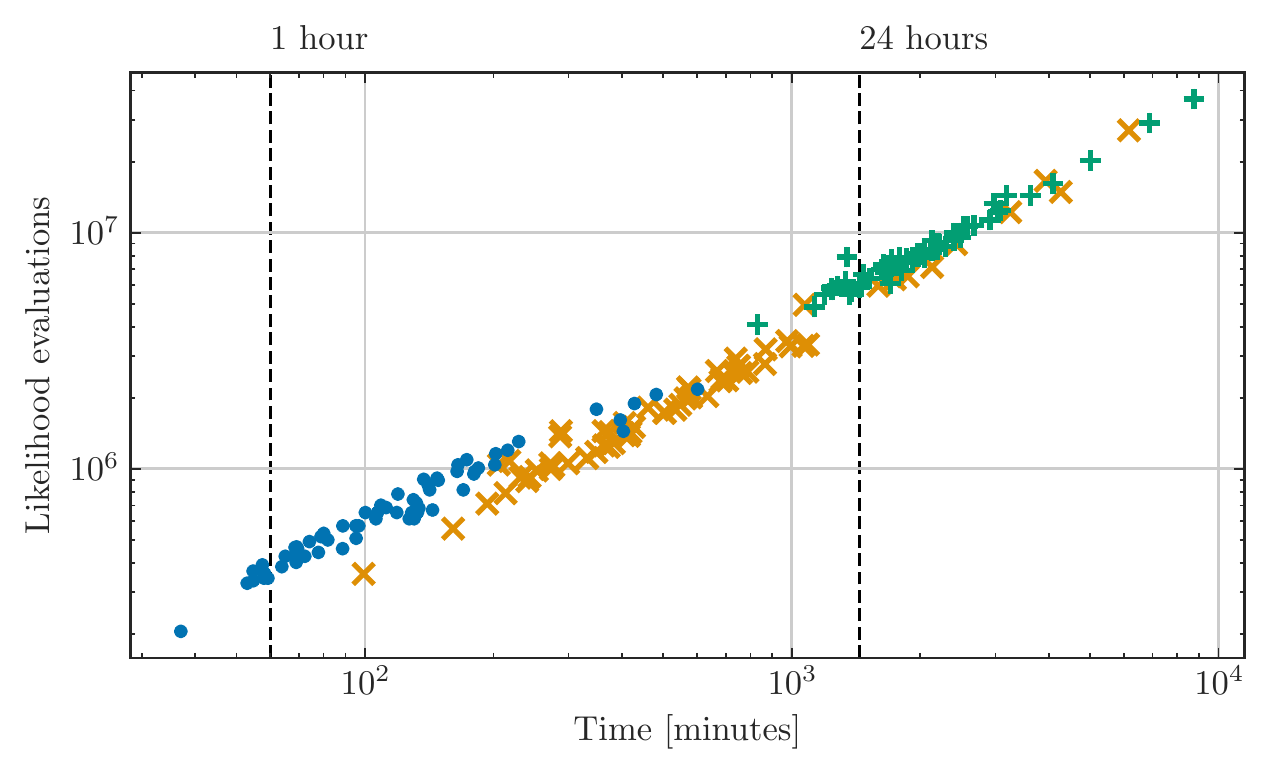}
    \end{indented}
    \caption{Total sampling time versus number of likelihood evaluations for \inessai (blue dots), \nessai (orange crosses) and \dynesty (green plus signs)  for the \numbbhsignals binary black hole injections described in \cref{sec:bbh_results}.}
    \label{fig:bbh_comp}
\end{figure}

In \cref{fig:bbh_comp}, we show the sampling time and the number of likelihood evaluations required to reach convergence. The median number of likelihood evaluations is \bbhmedianlikelihoodevals and the median wall time is \bbhmediantime. We also include results obtained using \nessai and \dynesty\cite{dynesty}\footnote{We use \dynesty version 1.0.1 with the custom random walk implementation included in \bilby version 1.2.1~\cite{bilby,bilby-gwtc}}, which has been used extensively for gravitational-wave inference \cite{gwtc2,gwtc2.1,gwtc3,bilby-gwtc}. Probability-probability plots for both samplers are shown in \cref{app:pp_plots}. We observe that the median reduction in the number of likelihood evaluations are \bbhlikelihoodratio and \bbhlikelihoodratiodynesty for \nessai and \dynesty respectively. These equate to reductions in the total wall time of \bbhtimeratio times and \bbhtimeratiodynesty times.

\subsection{Binary neutron star analysis with reduced order quadrature bases}\label{sec:bns}

We simulate the signal from a binary neutron star merger similar to GW190425~\cite{gw190425} at a distance of \qty{\bnsdistance}{Mpc} using \imrphenomptidal~\cite{imrphenompv2_nrtidalv2} and inject it into \bnsduration seconds of simulated noise from a two-detector network with aLIGO noise spectral density sensitivity \cite{LIGOScientific:2014pky} sampled at \qty{\bnssamplingfrequency}{Hz}. The resulting signal has an optimal network SNR of \bnsnetworksnr. 

To analyse the signal, we use \imrphenomp~\cite{Hannam:2013oca,Husa:2015iqa,Khan:2015jqa} with a \gls{roq} basis~\cite{Smith:2016qas} to reduce the cost of evaluating the likelihood\footnote{We use the \gls{roq} data available at \url{https://git.ligo.org/lscsoft/ROQ_data}.}. We also limit the analysis to assume aligned spins and use a low-spin prior a described in \citeauthor{gw190425}~\cite{gw190425}. We run the analysis using \inessai, \nessai and \dynesty. We repeat each analysis with four different random seeds and combine the posterior distributions for each seed into a single distribution. We use 16 cores for each analysis to decrease the overall wall time. The settings for \inessai are tuned to ensure that the effective number of posterior samples are comparable to the other samplers.

\begin{figure}
    \begin{indented}\item[]
        \centering
        \includegraphics{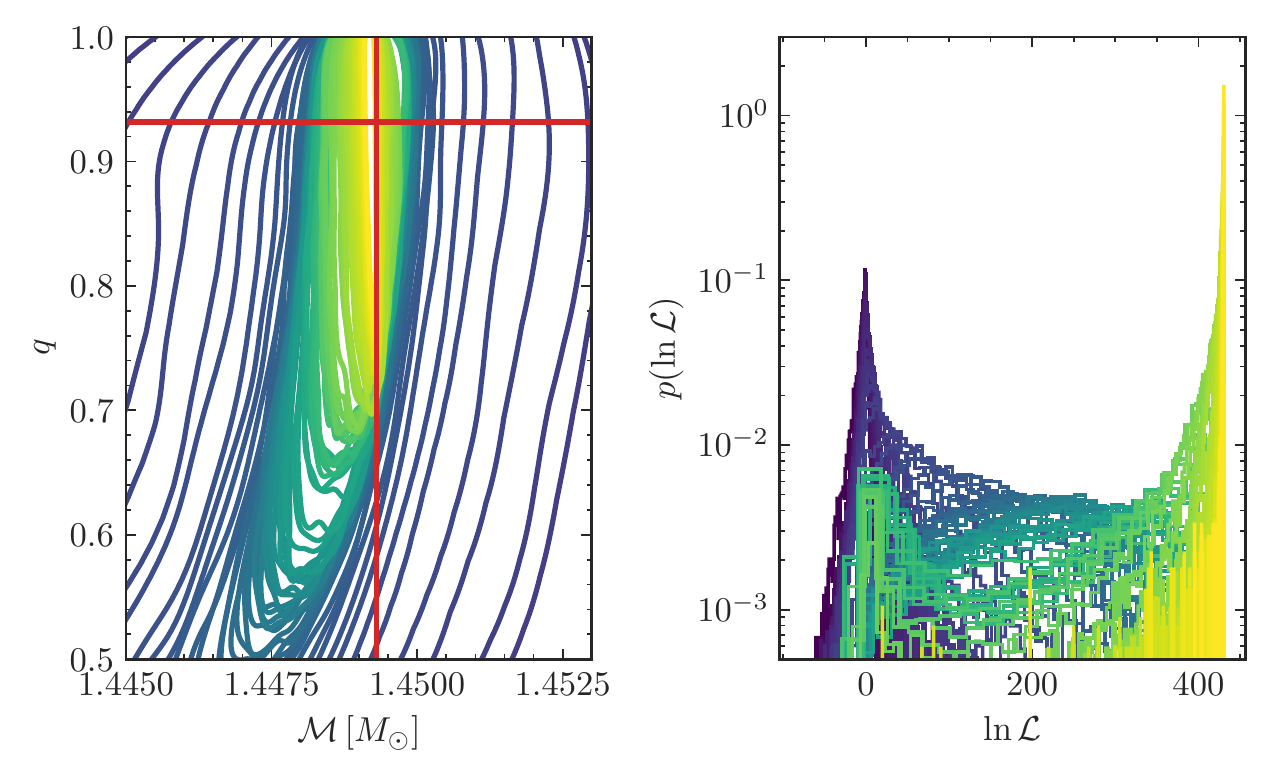}
    \end{indented}
    \caption{Evolution of the proposal distributions ($q_i(\theta)$) included in the meta-proposal when performing inference on the binary neutron star injection described in \cref{sec:bns}. Brighter colours indicate later iterations in the algorithm. \textbf{Left:} the 90\% contours for each of the proposal distributions in the chirp mass-mass ratio space. Only a small region of the parameter space around the highest likelihood is shown. The cross-hair indicates the injected value. \textbf{Right:} the distribution of log-likelihoods for each of the proposal distributions.}
    \label{fig:bns_likelihood_levels}
\end{figure}

In \cref{fig:bns_likelihood_levels}, we show how the meta-proposal evolves as more proposal distributions (normalising flows) are added over the course of sampling. This shows how the proposals converge around the parameters of the injected signal which correspond to the region with the highest log-likelihood.

To quantify the differences between the results, we compute the \gls{jsd} between the marginal posterior distributions for each parameter as described in \citeauthor{bilby-gwtc}~\cite{bilby-gwtc}. We use the threshold described in \citeauthor{bilby_mcmc}~\cite{bilby_mcmc} to determine if the \gls{jsd} indicate significant statistical differences between the results. We find that all the divergences are below the threshold, except for the in-plane spin $\chi_1$, for which \inessai and \nessai agree but \dynesty marginally disagrees with both. We include the complete set of \glspl{jsd} in \cref{app:jsd} and a corner plot comparing the distributions in \cref{app:bns_corner}.

We also compare the total number of likelihood evaluations and wall time for each sampler in \cref{tab:bns_results}. From these results we see that, on average, \inessai requires \inessaievalsnessai and \inessaievalsdynesty times fewer likelihood evaluations than \nessai and \dynesty respectively.

\begin{table}
    \caption{Total likelihood evaluations, wall time in minutes and \gls{ess} of the posterior distribution for the binary neutron star analysis with \glspl{roq} as described in \cref{sec:bns} for \dynesty, \nessai and \inessai. Results are averaged over four runs and the mean and standard deviations are quoted. All analyses were run with 16 cores.}
    \label{tab:bns_results}
    
    \begin{indented}
    \lineup
    \item[]\begin{tabular}{lccc}
\br
 & Wall time [min] & Likelihood evaluations & Effective sample size \\
\mr
\codestyle{dynesty} & $376.3 \pm 8.1$ & $\num{4.30e+07}\pm\num{7.12e+04}$ & $13098 \pm 131$ \\
\codestyle{nessai} & $57.9 \pm 8.9$ & $\num{1.42e+06}\pm\num{1.74e+05}$ & $13036 \pm 45$ \\
\codestyle{i-nessai} & $24.3 \pm 3.0$ & $\num{1.01e+06}\pm\num{8.99e+04}$ & $14625 \pm 3539$ \\
\br
\end{tabular}

    \end{indented} 
\end{table}

\subsection{Parallelisation}

As mentioned previously, the formulation of nested sampling used in this work does not have the same serial limitations of standard nested sampling. The algorithm we present is designed around drawing new samples and evaluating their likelihood in parallel. This leverages the inherently parallelised nature of the normalising flows. However, the process of training subsequent proposals to add to the meta-proposal is still a serial process.

\begin{figure}
    \begin{indented}\item[]
        \centering
        \includegraphics[width=\figwidth]{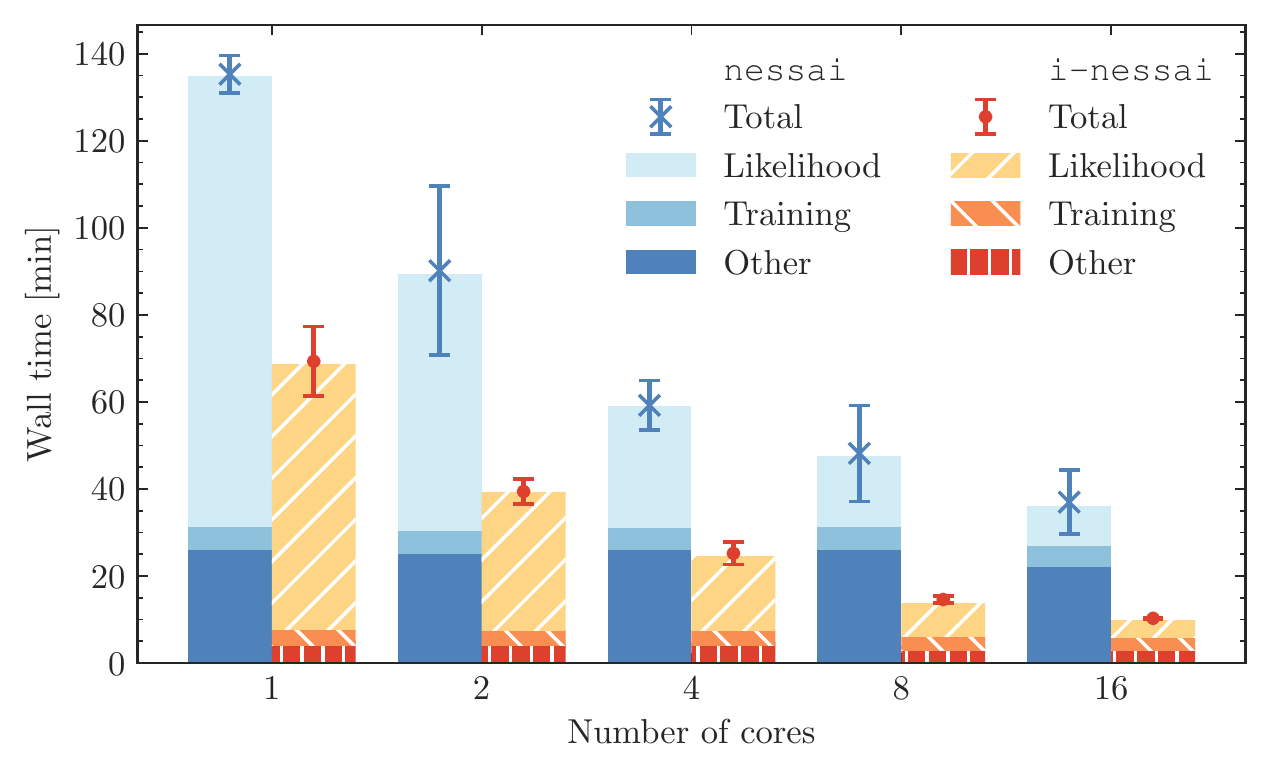}
    \end{indented}
    \caption{Comparison of the wall time spent training the normalising flows and evaluating the likelihood in \nessai and \inessai as a function of the number of cores. Results are shown for one of the binary black hole injections described in \cref{sec:bbh_results} and are averaged over four runs.}
    \label{fig:likelihood_parallelisation}
\end{figure}

In standard \nessai, the costs of rejection sampling and training set an upper limit for the reduction in wall time that can be achieved by parallelising the likelihood evaluation. However, the total cost of training typically accounted for less than \nessaitrainingtime of the total wall time~\cite{nessai}. In \inessai, the rejection sampling step is no longer necessary, so the training is now the main limiting factor and the potential reduction in wall time is far greater. In \cref{fig:likelihood_parallelisation}, we present results showing how the wall time decreases for an increasing number of cores for one of the binary black holes injections used in \cref{sec:bbh_results}. This shows how initially the wall time is dominated by the cost of evaluating the likelihood but as more cores are added the inherent cost of sampling, which includes training the flows and drawing new samples, becomes the dominant cost. However, in this example, it only accounts for $\inessaiothertime$ of the total wall time when running on a single core.

\subsection{Algorithm scaling}\label{sec:scaling}

In \inessai the number of live points has a different function to that in a typical nested sampler since, in combination with the method used to determine new levels, it will determine how many points are removed at an iteration and how many remain to train the normalising flow. We previously noted that, for \nessai, 2000 points were needed for reliable results~\cite{nessai}. We now test \inessai with different values of $\nlive$ and set the number of samples per flow $\nperflow = \nlive$

\begin{figure}
    \centering
    \includegraphics{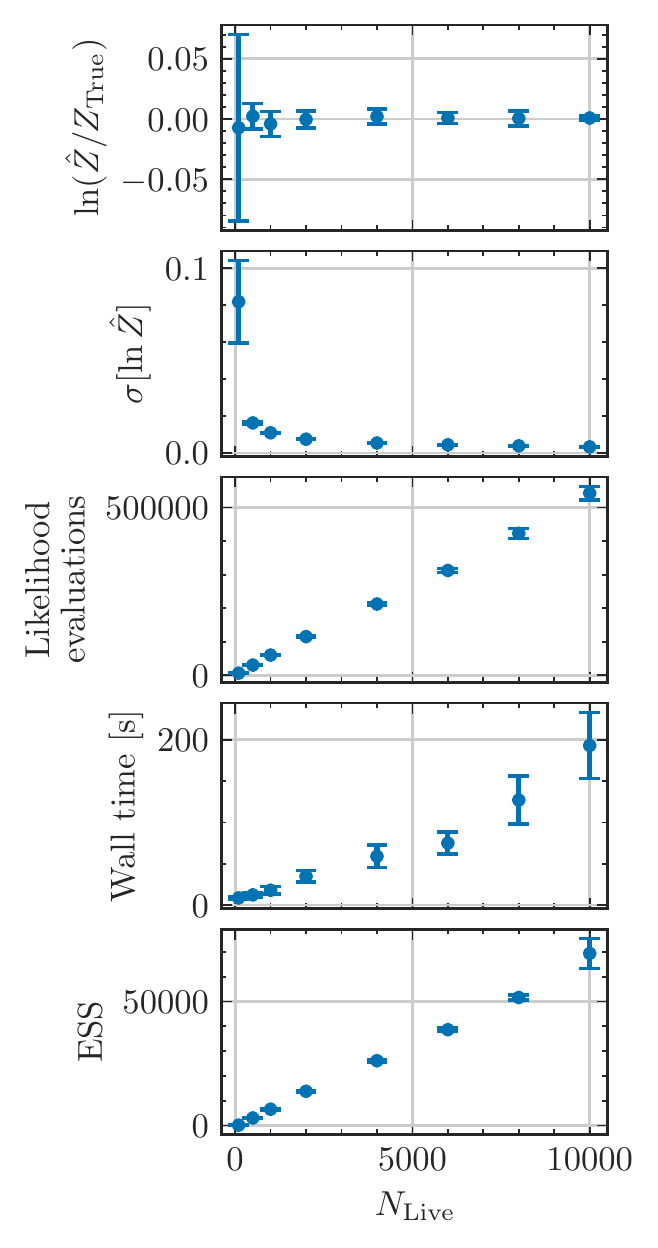}
    \caption{Scaling of \inessai as a function of the number of live points $\nlive$ for an \protect\ndimensional{16} Gaussian likelihood, as described in \cref{sec:analytic_likelihoods}. Results are averaged over 10 runs and the error-bars show the observed standard deviation. From top to bottom the results show the mean estimated log-evidence rescaled by the true value, the mean estimated standard deviation for the log-evidence, the total number of likelihood evaluations, the total wall time and the \glsfirst{ess} of the posterior distribution as defined in \cref{eq:ess}.}
    \label{fig:scaling}
\end{figure}

We evaluate the scaling of \inessai as a function of $\nlive$ and present the results in \cref{fig:scaling} for a \ndimensional{16} Gaussian likelihood sampled with $\nlive = \{100, 500, 1000, 2000, 4000, 6000, 8000, 10000\}$. The estimated log-evidence is consistent with the true value for all values of $\nlive$ and both the observed and estimated standard deviations decrease as $\nlive$ increases, which is consistent with \cref{eq:is_evidence,eq:ins_evidence_error}. We observe that the number of likelihood evaluations scales approximately linearly with the number of live points. This contrasts with the wall time which, for a 100 times increase in the number of live points, only increases by $\nlivetimeincrease$ times. This is the result of using a  likelihood that has a low computational cost, so the cost of running the sampler is dominated by the operations related to the normalising flow: training, drawing new samples and computing the meta-proposal probability as given by \cref{eq:meta_proposal}. In practice, most likelihoods will have a higher computational cost and the wall time will scale approximately linearly with $\nlive$.

\section{Discussion and conclusions}\label{sec:discussion}

In this work, we present an importance sampling-based nested sampling algorithm, \inessai, that builds on existing work \cite{diffusive-ns,multinest_importance,ns-smc} to incorporate normalising flows and overcome the main bottlenecks in \nessai described in \citeauthor{nessai}~\cite{nessai}. The resulting algorithm is a hybrid between standard nested sampling and \gls{smc}, where normalising flows are successively trained and added to an overall meta-proposal that describes the distribution of samples. 

We demonstrate that \inessai reliably estimates the log-evidence and associated error for Gaussian and Gaussian Mixture likelihoods in up to 32 dimensions. When we compare these results to those obtained with standard \nessai, we observe that \inessai converges significantly faster and requires fewer overall likelihood evaluations. Furthermore, the observed variance in the estimated log-evidence is consistently less than for \nessai. This demonstrates that \inessai produces consistent evidence estimates at a fraction of computational cost while also being more precise.

We perform inference on \numbbhsignals simulated gravitational-wave signals from binary black hole coalescence using \inessai and show that it passes a probability-probability test (\cref{fig:bbh_pp_plot}) which indicates that it produces unbiased estimates of the system parameters. Furthermore, these results are obtained without introducing problem specific reparameterisations. Similarly to the analytic likelihoods, we compare these results to those obtained with \nessai and \dynesty and observe a median reduction in the number of likelihood evaluations of \bbhlikelihoodratio and \bbhlikelihoodratiodynesty times respectively, which equates to a \bbhtimeratio and \bbhtimeratiodynesty times reduction in the total wall time.

To further demonstrate the advantages of \inessai compared to standard samplers, we perform inference on a simulated GW190425-like binary neutron star merger using \gls{roq} bases \cite{Smith:2016qas} and aligned low-spin priors. The inference completes in just \inessaibnstime minutes, \inessaibnsspeedup and \inessaibnsspeedupdynesty times faster than \nessai and \dynesty respectively, while also producing consistent posterior distributions and only requiring \inessaibnsevals likelihood evaluations compared to \nessaibnsevals and \dynestybnsevals respectively.

We also show how the likelihood evaluation can be parallelised in \inessai and find that, once of the cost of evaluating the likelihood becomes negligible, training the normalising flows and drawing new samples are the main limiting factors. This is in contrast to \nessai, where performing rejection sampling is the main limiting factor, accounting for approximately \nessaipopulationtime of the time when running on a single core. In \inessai training and drawing new samples account for significantly less of the total time. It therefore has improved scaling with respect to the number of cores compared to \nessai, as shown in \cref{fig:likelihood_parallelisation}.

A downside of this approach when compared to \nessai is that the order statistics-based tests proposed in \citeauthor{Fowlie:2020mzs}~\cite{Fowlie:2020mzs} and included in \nessai are no longer applicable since we no longer require points be distributed according to the likelihood-constrained prior. It is therefore harder to identify under- or over-constraining in \inessai. The \gls{ess} (\cref{eq:ess}) can be used to diagnose issues during sampling, however it is not always a reliable diagnostic.

In future work we will consider alternative methods for constructing the meta-proposal which do not rely on discard samples, for example using only the weights in \cref{eq:weighted_kl} and we will explore optimising the meta-proposal weights after sampling. We will also explore applications of \inessai more complete gravitational-wave analyses like those described in \cite{gwtc2,gwtc2.1,gwtc3} which included calibration uncertainties and waveforms with higher-order modes. Another possible application to explore is model comparison; typically, if we want to obtain a posterior distribution for a different prior than that used for the sampling, the existing posterior samples must be re-weighted using an alternative prior. However, the formulation of the nested sampling in this work would allow for the prior to be changed post-sampling and the evidence recomputed by updating \cref{eq:evidence_sum}, so long as the new prior does not extend the boundaries of the prior using during the initial sampling.

In summary, we have introduced an importance nested sampling algorithm, \inessai, that leverages normalising flows and addresses the bottlenecks in \nessai~\cite{nessai}. We have demonstrated that \inessai produces results that are consistent with standard nested sampling for a range of problems, whilst requiring up to an order-of-magnitude fewer likelihood evaluations and having improved scalability. Similarly to \nessai, \inessai is a drop-in replacement for existing samplers, meaning it can easily be used to accelerate existing analyses.

\ack

The authors thank Jordan McGinn and Federico Stachurski for insightful discussions about training normalising flows with weights and Greg Ashton for providing the code for computing Jensen-Shannon divergences. The authors also thank the members of the Data Analysis Group of the Institute of Gravitational Research and the LVK Parameter Estimation group for helpful discussions. The authors thank the two anonymous referees for their suggestions, which helped improve the manuscript. The authors gratefully acknowledge the Science and Technology Facilities Council of the United Kingdom. M.J.W. is supported by the Science and Technology Facilities Council [2285031]. J.V. and C.M. are supported by the Science and Technology Research Council [ST/V005634/1]. M.J.W. and C.M. are also supported by the European Cooperation in Science and Technology (COST) action [CA17137]. The authors are grateful for computational resources provided by Cardiff University and the LIGO Laboratory, and funded by the STFC grant [ST/I006285/1] supporting UK Involvement in the Operation of Advanced LIGO and the National Science Foundation Grants PHY-0757058 and PHY-0823459 respectively.

\textit{Software:} \nessai is implemented in \python and uses \numpy~\cite{numpy}, \scipy~\cite{2020SciPy-NMeth}, \pandas~\cite{reback2020pandas,mckinney-proc-scipy-2010}, \nessaimodels~\cite{nessai_models}, \nflows~\cite{nflows}, \glasflow~\cite{glasflow}, \pytorch~\cite{Paszke:2019:pt}, \matplotlib~\cite{Hunter:2007} and \seaborn~\cite{waskom2020seaborn}. Gravitational-wave injections were generated and analysed using \lalsuite~\cite{lalsuite}, \bilby and \bilbypipe~\cite{bilby}. The analysis also made use of \statsmodels~\cite{statsmodels}. Figures were prepared using \matplotlib~\cite{Hunter:2007}, \seaborn~\cite{waskom2020seaborn}, and \corner~\cite{corner}.

\appendix

\section{Weighted Kullback-Leilber divergence}\label{app:weighted_kl}

The \gls{kld} of two distributions $p(x)$ and $q(x)$ is defined as
\begin{equation}
    \kl(p, q) = \int p(x) \log \left[ \frac{p(x)}{q(x)}\right] \diff x.
\end{equation}
If we consider the case of minimising the \gls{kld} between two distributions $p(x)$ and $q(x)$ where $p(x)$ is fixed, then
\begin{equation}
    \begin{split}
        \kl(p, q) = &- \int p(x) \log q(x) \diff x + \int p(x) \log p(x) \diff x, \\
                    & -\int p(x) \log q(x) \diff x + \textrm{constant}.        
    \end{split}
\end{equation}
The constant term is independent of $q(x)$ so we only need to compute the first term when minimising the \gls{kld}. Using a Monte Carlo approximation of the integral with samples $x$ drawn from $r(x)$ this becomes
\begin{equation}\label{eq:kl_approx}
    \kl(p, q) \approx \widehat{\kl}(p, q) = -\frac{1}{N} \sum_{i=1}^{N} \frac{p(x_{i})}{r(x_{i})}\log q(x_{i}) + \textrm{constant}.
\end{equation}
If $r \equiv p$ and this reduces to
\begin{equation}
    \widehat{\kl}(p, q) = -\frac{1}{N} \sum_{i=1}^{N}\log q(x_{i}) + \textrm{constant},
\end{equation}
and we can ignore the constant when optimising $q(x)$. However, if $r\not\equiv p$ and both $p(x)$ and $r(x)$ are tractable, then we can define
\begin{equation}
    \widehat{\kl}(p, q) = -\frac{1}{N} \sum_{i=1}^{N} w_i \log q(x_{i}) + \textrm{constant},
\end{equation}
where $w_i\equiv{p(x_i)}/{r(x_i)}$. This allows for the \gls{kld} to be minimised using samples that are not from the target distribution.

\section{Methods for constructing the next proposal distribution}\label{app:level}

We test the quantile-based method and the entropy-based methods for constructing the next proposal distribution described in \cref{sec:level_method} and consider the stability and number of iterations required to converge. We find that the quantile-based method for determining the next level is sensitive to outliers in the meta-proposal $Q(\theta)$. This leads to large changes in the number of discarded samples $\nremoved$ between iterations which in turn can make the algorithm unstable. In contrast, the entropy-based approach is far more stable and leads to smoother variations in the number of discarded samples which we attribute to the use of the log-weights. Additionally, we find that the entropy-based method converges quicker than the quantile-based because the prior volume shrinks faster. As such, we use the entropy-based method for all our experiments.

\clearpage

\section{Validating the variance estimator}\label{app:validating_variance}

\begin{figure}[h]
    \begin{indented}\item[]
        \centering
        \includegraphics{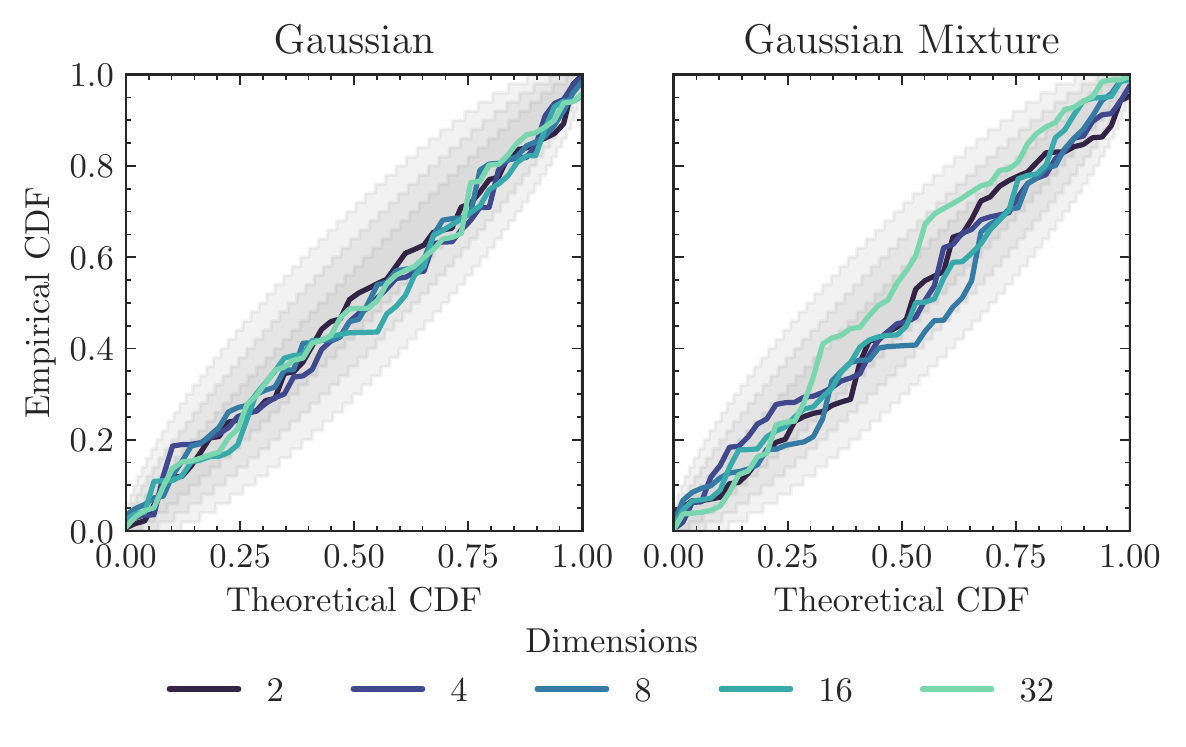}
    \end{indented}
    \caption{Probability-probability (P-P) plots for the estimated evidences for the Gaussian and Gaussian Mixture models described in \cref{sec:analytic_likelihoods} for $n=\{2, 4, 8, 16, 32\}$. The theoretical distribution is assumed to be a Gaussian centred around the true evidence with the standard deviation given by the estimated standard deviation \cref{eq:ins_evidence_error} averaged over \nanalytictests analyses per dimension. The 1-, 2- and 3-$\sigma$ confidence intervals are indicated by the shaded regions.}
    \label{fig:uncertainty_pp_plots}
\end{figure}

We validate the unbiased estimator for the variance of the evidence from \cref{eq:ins_evidence_error} for \inessai using the Gaussian and Gaussian Mixture likelihoods described in \cref{sec:analytic_likelihoods}. We use the results from the analyses described in \cref{sec:analytic_likelihoods} and produce probability-probability (P-P) plots comparing the observed distribution of evidences and a Gaussian distribution with the mean equal to the true evidence and the standard deviation estimated using \cref{eq:ins_evidence_error} averaged over the \nanalytictests runs per dimensions. The results are presented in \cref{fig:uncertainty_pp_plots} and show good agreement between the estimated and observed distributions.

\clearpage

\section{Insertion indices test for the Rosenbrock likelihood}\label{app:rosenbrock_p_values}

\begin{figure}[h]
    \begin{indented}\item[]
        \centering
        \includegraphics{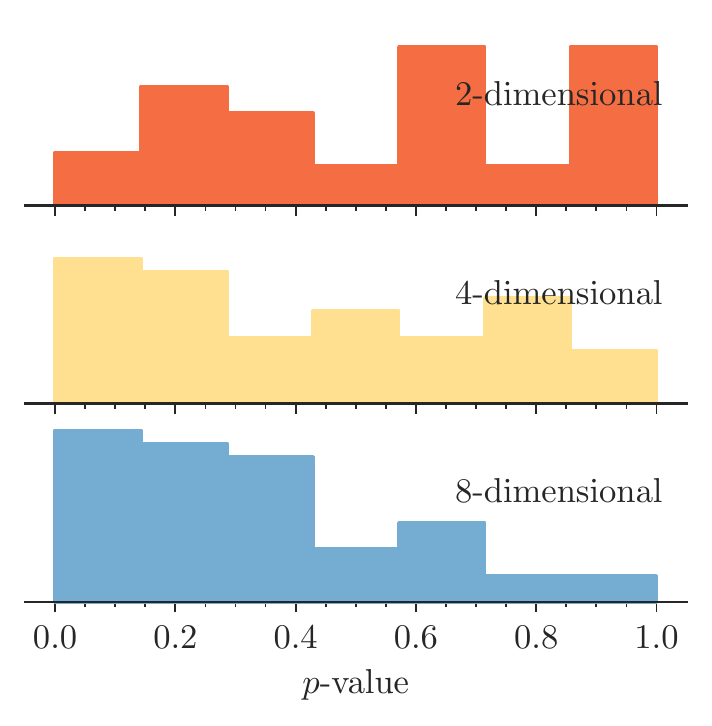}
    \end{indented}
    \caption{Distribution of the $p$-values for the insertion indices \cite{Fowlie:2020mzs} when analysing the Rosenbrock likelihood \nanalytictests times using \nessai with $n=\{2, 4, 8\}$. }
    \label{fig:rosenbrock_p_values}
\end{figure}

In \cref{sec:rosenbrock}, we analyse the Rosenbrock likelihood for  $n=\{2, 4, 8\}$ using \nessai and \inessai and find that the estimated log-evidence disagreed as shown in \cref{fig:comparison}. In \citeauthor{Fowlie:2020mzs}~\cite{Fowlie:2020mzs}, the authors proposed using order-statistics to check the convergence of nested sampling runs. This involves computing an insertion index for each new sample according to where it is inserted into the current ordered set of live points. If new samples are distributed according to the prior, then the overall distribution of the insertion indices should be uniform. This can be quantified by computing a $p$-value for the overall distribution using the Kolmogorov-Smirnov statistic~\cite{Smirnov:1948table} for discrete distributions~\cite{Arnold:2011npgof}. We compute $p$-values for each analysis and presented the results in \cref{fig:rosenbrock_p_values}. If the results are unbiased then the distribution of $p$-values should be uniform on $[0, 1]$, however we observe that for $n>2$ the distributions are not uniform, indicating problems during sampling. This agrees with the observation that for $n=\{4, 8\}$, with the settings used, \nessai over-estimates the log-evidence.

\clearpage

\section{Probability-probability plots for other samplers}\label{app:pp_plots}

\begin{figure}[h]
    \begin{indented}\item[]
        \centering
        \subfigure[\nessai]{\includegraphics[width=0.65\linewidth]{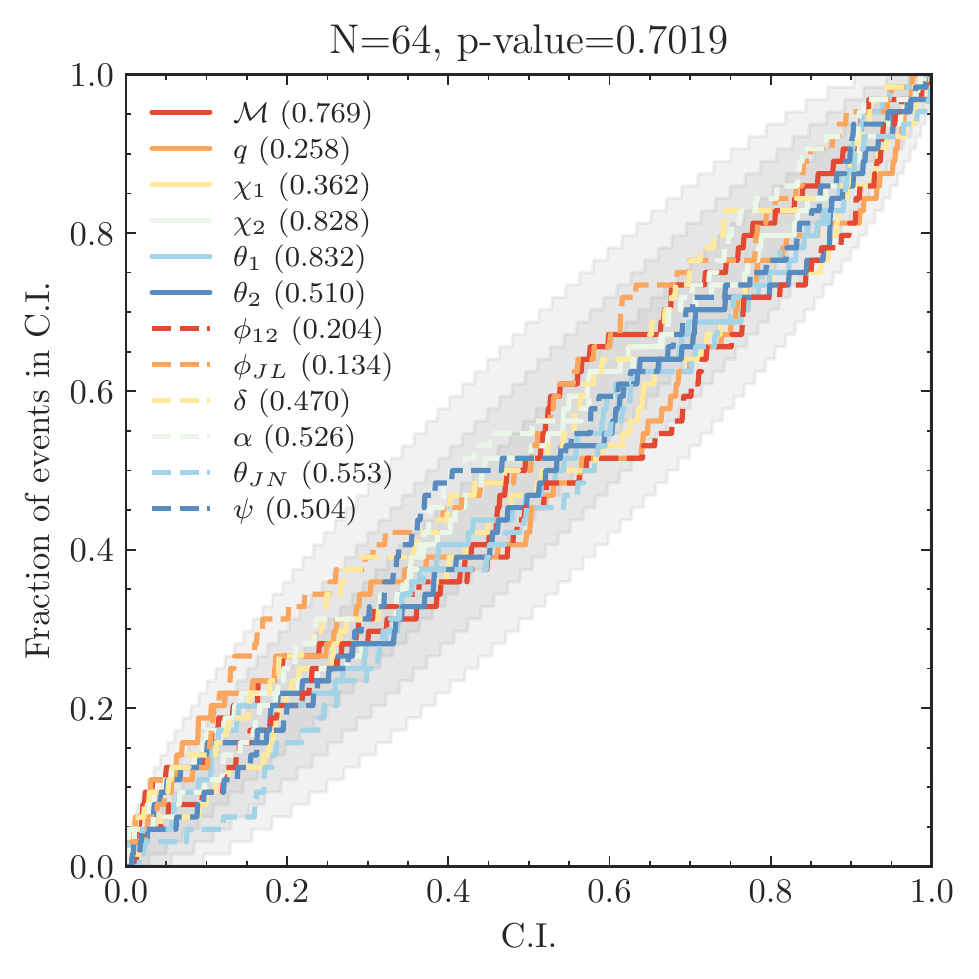}}
        \subfigure[\dynesty]{\includegraphics[width=0.65\linewidth]{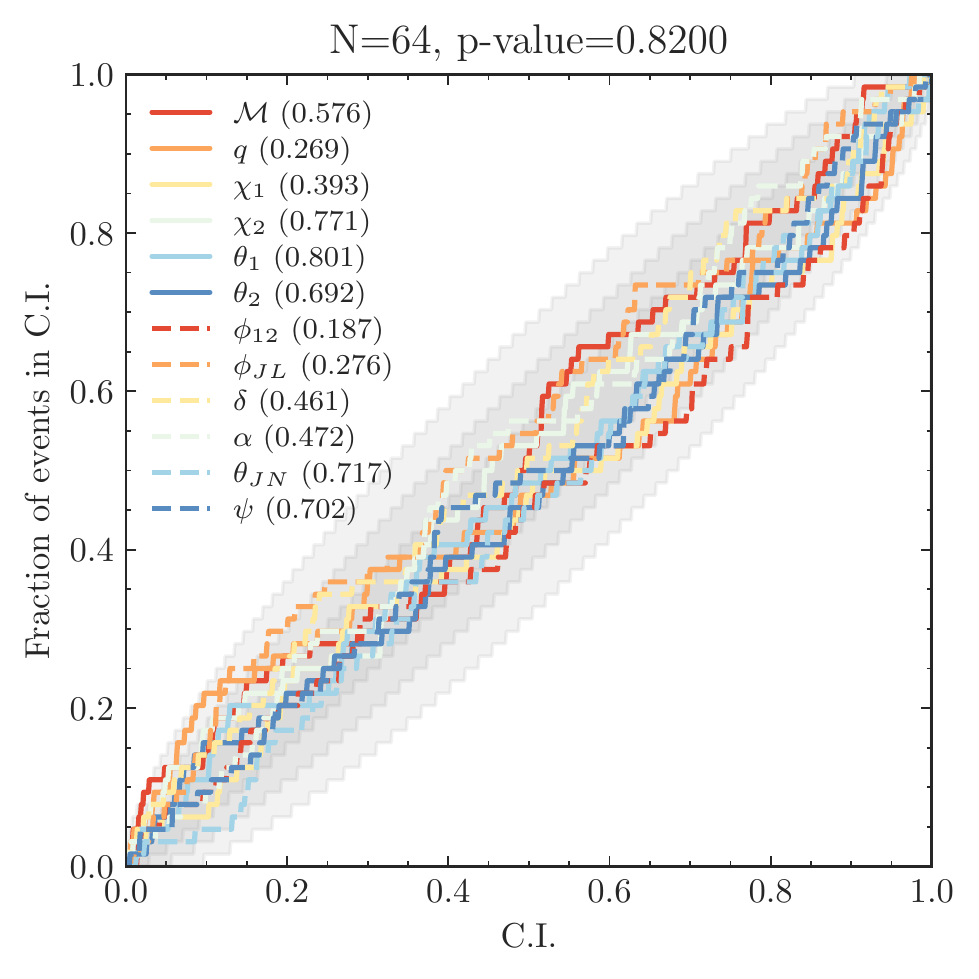}}
    \end{indented}
    \caption{Probability-probability (P-P) plot showing the confidence interval versus the fraction of the events within that confidence interval for the posterior distributions obtained using \nessai and \dynesty for \numbbhsignals simulated compact binary coalescence signals produced with \bilby and \bilbypipe. The 1-, 2- and 3-$\sigma$ confidence intervals are indicated by the shaded regions and $p$-values are shown for each of the parameters and the combined $p$-value is also shown.}
    \label{fig:pp_plot_other}
\end{figure}

\clearpage

\section{Jensen-Shannon divergence for comparing marginal posterior distributions}\label{app:jsd}

We compute the \gls{jsd} between the marginal posterior distributions obtained in \cref{sec:bns} as described in \citeauthor{bilby-gwtc}\cite{bilby-gwtc}. We use bootstrapping to compare 100 different realisations of \jsdsamples samples from each posterior and quote the mean \gls{jsd} and standard deviation in \cref{tab:jsd}. Following \citeauthor{bilby_mcmc}~\cite{bilby_mcmc}, for \jsdsamples posterior samples, the \gls{jsd} threshold is \qty{2e-3}{nats}. The divergences between \inessai and \nessai agree for all the parameters, whereas for \dynesty there is marginal disagreement in the posteriors for the aligned spin $\chi_1$. However, since \nessai and \inessai are in agreement, we do not investigate this further in this work.

\begin{table}[h]
    \caption{Jensen-Shannon divergences in units of \qty{1e-3}{nats} for the marginal posterior distributions between \nessai, \inessai and \dynesty. Values shown are the mean and the 1-$\sigma$ quantiles computed over 100 different realisations of \jsdsamples samples.}
    \label{tab:jsd}
    
    \begin{indented}
    \lineup
    \item[]\begin{tabular}{llll}
\br
 & \dynesty-\nessai & \dynesty-\inessai & \nessai-\inessai \\
\mr
$\mathcal{M}$ & $0.61^{0.20}_{-0.20}$ & $0.69^{0.22}_{-0.19}$ & $0.53^{0.21}_{-0.13}$ \\
$q$ & $0.52^{0.29}_{-0.16}$ & $0.36^{0.22}_{-0.11}$ & $0.30^{0.18}_{-0.08}$ \\
$\chi_1$ & $2.24^{0.78}_{-0.55}$ & $2.61^{0.77}_{-0.59}$ & $0.53^{0.27}_{-0.17}$ \\
$\chi_2$ & $1.68^{0.60}_{-0.46}$ & $1.93^{0.47}_{-0.54}$ & $0.73^{0.22}_{-0.22}$ \\
$\delta$ & $1.37^{0.29}_{-0.28}$ & $1.47^{0.34}_{-0.28}$ & $1.59^{0.38}_{-0.31}$ \\
$\alpha$ & $1.04^{0.22}_{-0.25}$ & $1.15^{0.25}_{-0.27}$ & $1.37^{0.30}_{-0.28}$ \\
$\theta_{JN}$ & $0.71^{0.22}_{-0.17}$ & $0.74^{0.21}_{-0.21}$ & $0.79^{0.26}_{-0.23}$ \\
$\psi$ & $0.18^{0.13}_{-0.06}$ & $0.21^{0.15}_{-0.09}$ & $0.19^{0.10}_{-0.09}$ \\
$t_\textrm{c}$ & $1.29^{0.42}_{-0.25}$ & $1.56^{0.31}_{-0.39}$ & $1.57^{0.39}_{-0.33}$ \\
\br
\end{tabular}

    \end{indented}
\end{table}

\clearpage

\section{Binary neutron star corner plot}\label{app:bns_corner}

\begin{figure}[h]
        \centering
        \includegraphics[width=\textwidth]{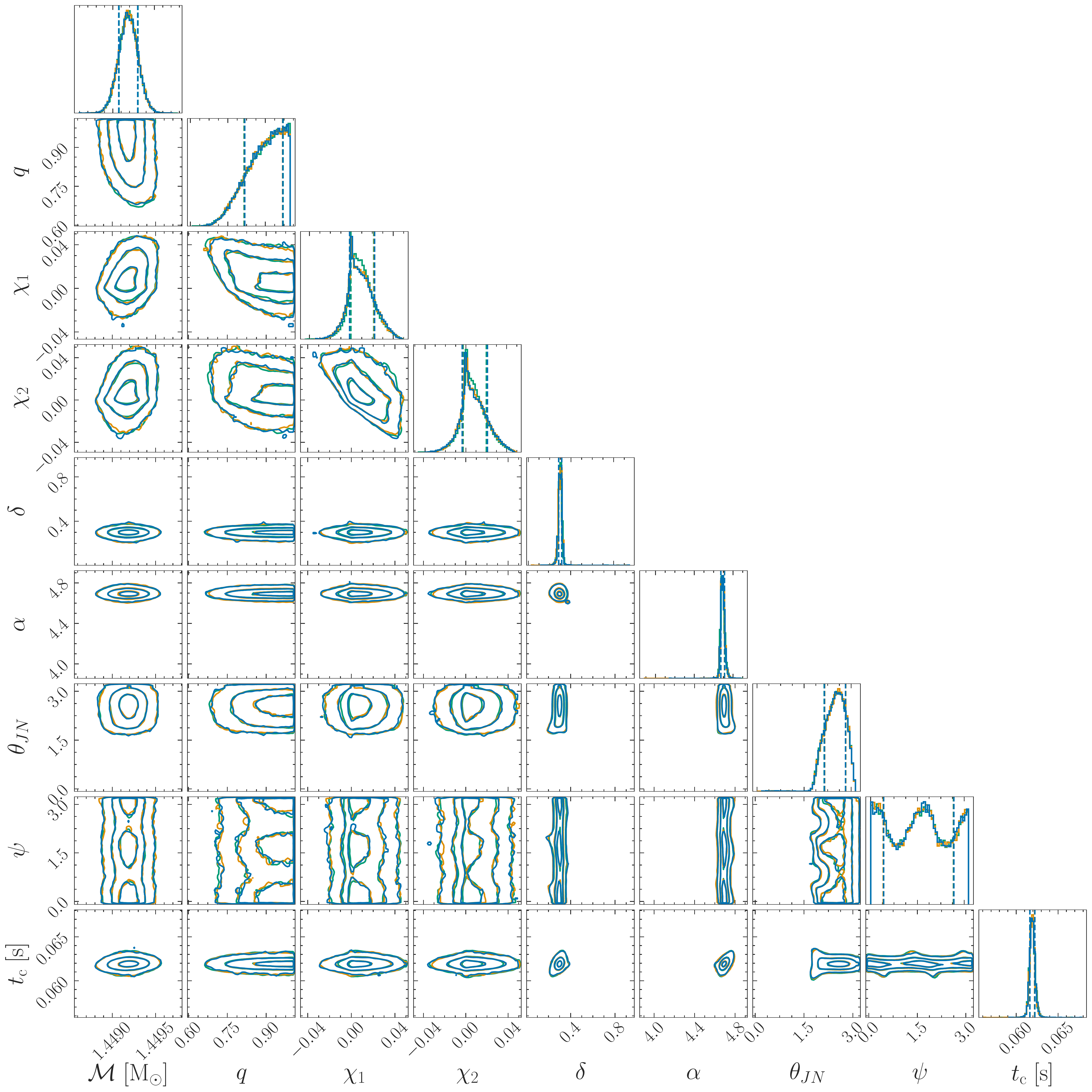}
    \caption{Posterior distributions for the GW190425-like injection described in \cref{sec:bns}. Results are shown for \dynesty in green, \nessai in orange and \inessai in blue. The 1-$\sigma$ confidence intervals for each parameter are shown in the marginal histograms.}
    \label{fig:bns_corner}
\end{figure}

\clearpage

\printbibliography

\end{document}